\documentclass[longauth]{aa} 
\usepackage[colorlinks=true,citecolor=blue]{hyperref}
\usepackage{xcolor}
\usepackage{tikz}
\usepackage{placeins}

\definecolor{lime}{HTML}{A6CE39}
\DeclareRobustCommand{\orcidicon}{%
    \begin{tikzpicture}
    \draw[lime, fill=lime] (0,0) 
    circle [radius=0.16] 
    node[white] {{\fontfamily{qag}\selectfont \tiny ID}};
    \draw[white, fill=white] (-0.0625,0.095) 
    circle [radius=0.007];
    \end{tikzpicture}
    \hspace{-2mm}
}
\newcommand{\orcid}[1]{\href{https://orcid.org/#1}{\orcidicon}}
\usepackage{natbib}
\bibpunct{(}{)}{;}{a}{}{,} 
\usepackage{graphicx}
\usepackage[varg]{txfonts}
\begin{document}

   \title{Validation and atmospheric exploration of the sub-Neptune TOI-2136b around a nearby M3 dwarf}


   \author{K. Kawauchi\inst{\ref{inst1}}\fnmsep\inst{\ref{inst2}}\orcid{0000-0003-1205-5108}
          \and
          F. Murgas\inst{\ref{inst1}}\fnmsep\inst{\ref{inst2}}\orcid{0000-0001-9087-1245}
          \and
          E. Palle\inst{\ref{inst1}}\fnmsep\inst{\ref{inst2}}\orcid{0000-0003-0987-1593}
          \and
          N. Narita\inst{\ref{inst3}}\fnmsep\inst{\ref{inst4}}\fnmsep\inst{\ref{inst5}}\fnmsep\inst{\ref{inst1}}\orcid{0000-0001-8511-2981}
          \and
          A. Fukui\inst{\ref{inst3}}\fnmsep\inst{\ref{inst1}}\orcid{0000-0002-4909-5763}
          \and
          T. Hirano\inst{\ref{inst5}}\fnmsep\inst{\ref{inst6}}\orcid{0000-0003-3618-7535}
          \and
          H. Parviainen\inst{\ref{inst1}}\fnmsep\inst{\ref{inst2}}\orcid{0000-0001-5519-1391}
          \and
          H. T. Ishikawa\inst{\ref{inst5}}\fnmsep\inst{\ref{inst6}}\orcid{00000-0001-6309-4380}
          \and
          N. Watanabe\inst{\ref{inst7}}\orcid{0000-0002-7522-8195}
          \and
          E. Esparaza-Borges\inst{\ref{inst1}}\fnmsep\inst{\ref{inst2}}\orcid{0000-0002-2341-3233}
          \and
          M. Kuzuhara\inst{\ref{inst5}}\fnmsep\inst{\ref{inst6}}\orcid{0000-0002-4677-9182}
          \and
          J. Orell-Miquel\inst{\ref{inst1}}\fnmsep\inst{\ref{inst2}}\orcid{0000-0003-2066-8959}
          \and
          V. Krishnamurthy\inst{\ref{inst5}}\fnmsep\inst{\ref{inst6}}\orcid{0000-0003-2310-9415}
          \and
          M. Mori\inst{\ref{inst8}}\orcid{0000-0003-1368-6593}
          \and
          T. Kagetani\inst{\ref{inst7}}\orcid{0000-0002-5331-6637}
          \and
          Y. Zou\inst{\ref{inst7}}\orcid{0000-0002-5609-4427}
          \and
          K. Isogai\inst{\ref{inst9}}\fnmsep\inst{\ref{inst7}}\orcid{0000-0002-6480-3799}
          \and
          J. H. Livingston\inst{\ref{inst10}}\fnmsep\inst{\ref{inst5}}\fnmsep\inst{\ref{inst6}}\orcid{0000-0002-4881-3620}
          \and
          S. B. Howell\inst{\ref{inst21}}\orcid{0000-0002-2532-2853}
          \and
          N. Crouzet\inst{\ref{inst13}}\orcid{0000-0001-7866-8738}
          \and
          J. P. de Leon\inst{\ref{inst8}}\orcid{0000-0002-6424-3410}
          \and
          T. Kimura\inst{\ref{inst25}}
          \and
          T. Kodama\inst{\ref{inst3}}\orcid{0000-0001-9032-5826}
          \and
          J. Korth\inst{\ref{inst27}}\orcid{0000-0002-0076-6239}
          \and
          S. Kurita\inst{\ref{inst25}}
          \and
          A. Laza-Ramos\inst{\ref{inst1}}\orcid{0000-0003-3316-3044}
          \and
          R. Luque\inst{\ref{inst26}}\orcid{0000-0002-4671-2957}
          \and
          A. Madrigal-Aguado\inst{\ref{inst1}}\fnmsep\inst{\ref{inst2}}\orcid{0000-0002-9510-0893}
          \and
          K. Miyakawa\inst{\ref{inst14}}\orcid{0000-0002-5706-3497}
          \and
          G. Morello\inst{\ref{inst1}}\fnmsep\inst{\ref{inst2}}\orcid{0000-0002-4262-5661}
          \and
          T. Nishiumi\inst{\ref{inst19}}\fnmsep\inst{\ref{inst5}}\fnmsep\inst{\ref{inst7}}\orcid{0000-0003-1510-8981}
          \and
          G. E. F. Rodr\'{i}guez\inst{\ref{inst1}}\orcid{0000-0003-0597-7809}
          \and
          M. S\'{a}nchez-Benavente\inst{\ref{inst1}}\fnmsep\inst{\ref{inst2}}\orcid{0000-0003-2693-279X}
          \and
          M. Stangret\inst{\ref{inst1}}\fnmsep\inst{\ref{inst2}}\orcid{0000-0002-1812-8024}
          \and
          H. Teng\inst{\ref{inst14}}\orcid{0000-0003-3860-6297}
          \and
          Y. Terada\inst{\ref{inst11}}\fnmsep\inst{\ref{inst12}}\orcid{0000-0003-2887-6381}
          \and
          C. L. Gnilka\inst{\ref{inst21}}\fnmsep\inst{\ref{inst22}}\orcid{0000-0003-2519-6161}
          \and
          N. Guerrero\inst{\ref{inst23}}\fnmsep\inst{\ref{inst24}}\orcid{0000-0002-5169-9427}
          \and
          H. Harakawa\inst{\ref{inst16}}\orcid{0000-0002-7972-0216}
          \and
          K. Hodapp\inst{\ref{inst17}}\orcid{0000-0003-0786-2140}
          \and
          Y. Hori\inst{\ref{inst5}}\fnmsep\inst{\ref{inst6}}\orcid{0000-0003-4676-0251}
          \and  
          M. Ikoma\inst{\ref{inst15}}\orcid{0000-0002-5658-5971}
          \and
          S. Jacobson\inst{\ref{inst17}}
          \and
          M. Konishi\inst{\ref{inst18}}\orcid{0000-0003-0114-0542}
          \and
          T. Kotani\inst{\ref{inst5}}\fnmsep\inst{\ref{inst6}}\fnmsep\inst{\ref{inst19}}
          \and
          T. Kudo\inst{\ref{inst16}}\orcid{0000-0001-6181-3142}
          \and
          T. Kurokowa\inst{\ref{inst5}}\fnmsep\inst{\ref{inst20}}
          \and
          N. Kusakabe\inst{\ref{inst5}}\fnmsep\inst{\ref{inst6}}\orcid{0000-0001-9194-1268}
          \and
          J. Nishikawa\inst{\ref{inst6}}\fnmsep\inst{\ref{inst19}}\fnmsep\inst{\ref{inst5}}\orcid{0000-0001-9326-8134}
          \and
          M. Omiya\inst{\ref{inst5}}\fnmsep\inst{\ref{inst6}}\orcid{0000-0002-5051-6027}
          \and
          T. Serizawa\inst{\ref{inst20}}\fnmsep\inst{\ref{inst6}}
          \and
          M. Tamura\inst{\ref{inst10}}\fnmsep\inst{\ref{inst5}}\fnmsep\inst{\ref{inst6}}\orcid{0000-0002-6510-0681}
          \and
          A. Ueda\inst{\ref{inst5}}\fnmsep\inst{\ref{inst6}}\fnmsep\inst{\ref{inst19}}
          \and
          S. Vievard\inst{\ref{inst16}}\fnmsep\inst{\ref{inst5}}
          }

   \institute{Instituto de Astrof\'\i sica de Canarias (IAC), 38205 La Laguna, Tenerife, Spain\label{inst1}
            \and
            Departamento de Astrof\'{i}sica, Universidad de La Laguna (ULL), 38206 La Laguna, Tenerife, Spain\label{inst2}
            \and
            Komaba Institute for Science, The University of Tokyo, 3-8-1 Komaba, Meguro, Tokyo 153-8902, Japan\label{inst3}
            \and
            JST, PRESTO, 3-8-1 Komaba, Meguro, Tokyo 153-8902, Japan\label{inst4}
            \and
            Astrobiology Center, 2-21-1 Osawa, Mitaka, Tokyo 181-8588, Japan\label{inst5}
            \and
            National Astronomical Observatory of Japan, 2-21-1 Osawa, Mitaka, Tokyo 181-8588, Japan\label{inst6}
            \and
            Department of Multi-Disciplinary Sciences, Graduate School of Arts and Sciences, The University of Tokyo, 3-8-1 Komaba, Meguro, Tokyo 153-8902, Japan\label{inst7}
            \and
            Department of Astronomy, Graduate School of Science, The University of Tokyo, 7-3-1 Hongo, Bunkyo-ku, Tokyo 113-0033, Japan\label{inst8}
            \and
            Okayama Observatory, Kyoto University, 3037-5 Honjo, Kamogatacho, Asakuchi, Okayama 719-0232, Japan\label{inst9}
            \and
            Department of Astronomy, The University of Tokyo, 7-3-1 Hongo, Bunkyo-ku, Tokyo 113-0033, Japan\label{inst10}
            \and
            NASA Ames Research Center, Moffett Field, CA 94035, US\label{inst21}
            \and
            European Space Agency (ESA), European Space Research and Technology Centre (ESTEC), Keplerlaan 1, 2201 AZ Noordwijk, The Netherlands\label{inst13}
            \and
            Department of Earth and Planetary Science, Graduate School of Science, The University of Tokyo, 7-3-1 Hongo, Bunkyo-ku, Tokyo 113-0033, Japan\label{inst25}
            \and
            Department of Space, Earth and Environment, Astronomy and Plasma Physics, Chalmers University of Technology, 412 96 Gothenburg, Sweden\label{inst27}
            \and
            Instituto de Astrof\'isica de Andaluc\'ia (IAA-CSIC), Glorieta de la Astronom\'ia s/n, 18008 Granada, Spain\label{inst26}
            \and
            Department of Earth and Planetary Sciences, Tokyo Institute of Technology, Meguro-ku, Tokyo, 152-8551, Japan\label{inst14}
            \and
            Department of Astronomy, School of Science, The Graduate University for Advanced Studies (SOKENDAI), 2-21-1 Osawa, Mitaka, Tokyo, Japan\label{inst19}
            \and
            Institute of Astronomy and Astrophysics, Academia Sinica, P.O. Box 23-141, Taipei 10617, Taiwan, R.O.C.\label{inst11}
            \and
            Department of Astrophysics, National Taiwan University, Taipei 10617, Taiwan, R.O.C.\label{inst12}
            \and
            NASA Exoplanet Science Institute, Caltech/IPAC, Mail Code 100-22, 1200 E. California Blvd., Pasadena, CA 91125, USA\label{inst22}
            \and
            Department of Astronomy, University of Florida, Gainesville, FL, 32611, USA\label{inst23}
            \and
            Department of Physics and Kavli Institute for Astrophysics and Space Research, Massachusetts Institute of Technology, Cambridge, MA 02139, USA\label{inst24}
            \and
            Subaru Telescope, 650 N. Aohoku Place, Hilo, HI 96720, USA\label{inst16}
            \and
            University of Hawaii, Institute for Astronomy, 640 N. Aohoku Place, Hilo, HI 96720, USA\label{inst17}
            \and
            Division of Science, National Astronomical Observatory of Japan, 2-21-1 Osawa, Mitaka, Tokyo 181-8588, Japan\label{inst15}
            \and
            Faculty of Science and Technology, Oita University, 700 Dannoharu, Oita 870-1192, Japan\label{inst18}
            \and
            Institute of Engineering, Tokyo University of Agriculture and Technology, 2-24-16, Naka-cho, Koganei, Tokyo, 184-8588, Japan\label{inst20}
            }

   \date{Received February 21, 2022; accepted June 14, 2022}


  \abstract
   {The NASA space telescope \textit{TESS} is currently in the extended mission of its all-sky search for new transiting planets. 
   Of the thousands of candidates that TESS is expected to deliver, transiting planets orbiting nearby M dwarfs are particularly interesting targets since they provide a great opportunity to characterize their atmospheres by transmission spectroscopy.}
   {We aim to validate and characterize the new sub-Neptune-sized planet candidate TOI-2136.01 orbiting a nearby M dwarf ($d = 33.36 \pm 0.02$ pc, $T_{eff} = 3373 \pm 108$ K) with an orbital period of 7.852 days.}
   {We use TESS data, ground-based multicolor photometry, and radial velocity measurements with the InfraRed Doppler (IRD) instrument on the Subaru Telescope to validate the planetary nature of TOI-2136.01, and estimate the stellar and planetary parameters. We also conduct high-resolution transmission spectroscopy to search for helium in its atmosphere.}
   {We confirm that TOI-2136.01 (now named TOI-2136b) is a bona fide planet with a planetary radius of $R_p = 2.20 \pm 0.07$ $\rm R_\oplus$ and a mass of $M_p = 4.7^{+3.1}_{-2.6}$ $\rm M_\oplus$. 
   We also search for helium 10830 \mbox{\AA} absorption lines and place an upper limit on the equivalent width of $<$ 7.8 m\mbox{\AA} and on the absorption signal of $<$ 1.44 \% with 95\% confidence.}
   {TOI-2136b is a sub-Neptune transiting a nearby and bright star (J=10.8 mag), and is a potentially hycean planet, which is a new class of habitable planets with large oceans under a H$_2$-rich atmosphere, making it an excellent target for atmospheric studies to understand the formation, evolution, and habitability of the small planets.}


   \maketitle
%

\section{Introduction}

    Small planets with radii between 1 - 4 $\rm R_\oplus$ are extremely common in the Milky Way, but do not exist in our solar system. NASA's Kepler space telescope revealed that this population has a bimodal distribution that separates the planets into super-Earth (1 - 1.5 $\rm R_\oplus$) and sub-Neptune (2 - 4 $\rm R_\oplus$) populations \citep{Fulton+2017} for planets with $P < 100$ days. This "radius valley" is consistent with the transition from rocky to non-rocky planets and is predicted by photoevaporation models \citep[e.g.,][]{Owen&Wu2013,Owen&Wu2017,Jin+2014,Lopez&Rice2018,Mordasini2020} and/or core-powered mass-loss models \citep{Ginzburg+2016,Ginzburg+2018,Gupta&Schlichting2019,Gupta&Schlichting2021}.

    Furthermore, the location of the radius valley is revealed to be dependent on the planetary period, insolation, and stellar type \citep[e.g.,][]{VanEylen+2018,Martinez+2019,Wu2019}. 
    The position of the radius gap in the radius-period distribution plane around Sun-like stars decreases with the orbital period following a power law $R_{\mathrm{gap}} \propto P^{-0.11}$, which is consistent with photoevaporation and core-powered mass-loss model predictions \citep{Martinez+2019}. On the other hand, it has been claimed that the position of the radius valley around late-type stars increases with the orbital period following a power law $R_{\mathrm{gap}} \propto P^{+0.058}$, which is consistent with gas-poor formation models \citep{Cloutier+2020}. However, a recent study found that the correlation between the position of the radius gap and orbital period is negative around M-type stars whose masses are above 0.5 $\rm M_{\odot}$ ($R_{\mathrm{gap}} \propto P^{-0.12}$), and this correlation does not seem to vary with the stellar type \citep{Petigura+2022}.
    To further explore the nature and origins of the radius valley around M dwarfs, it is important to discover new small planets and to measure their masses as well as radii, in order to have a more complete sample.

    Sub-Neptune-sized planets are likely to have H$_2$O-dominated ices and/or fluids (water worlds) in addition to a rocky core enveloped in a H$_2$-He gas \citep[e.g.,][]{Zeng2019}. Recent studies indicated the possibility that sub-Neptune-sized planets could have large oceans with habitable conditions under H$_2$-rich atmospheres \citep[Hycean worlds;][]{Madhusudhan+2021}.
    The direct observation of their atmospheres helps our understanding of the composition of the planets.

    The infrared helium triplet is a good indicator of a primary atmosphere accreted from the protoplanetary nebula \citep{Elkins-Tanton2008}. Previous studies have detected the \ion{He}{i} 10830\mbox{\AA} absorption lines in the atmosphere of about a half-dozen Jupiter- and Neptune-sized planets \citep[e.g.,][]{Spake+2018,Nortmann+2018,Allart+2018}, but only in three sub-Neptunes: GJ3470b \citep{Palle+2020}, GJ1214b \citep{Orell-Miquel+2022}, and TOI 560.01 \citep{Zhang+2022}. Other studies of sub-Neptunes have been able to place only upper limits on \ion{He}{i} absorption, but they still help to constrain the mass-loss rate \citep[e.g.,][]{Krishnamurthy+2021}.

    Here we report the discovery of TOI-2136b, a sub-Neptune-sized planet around an M dwarf, using photometry from NASA's Transiting Exoplanet Survey Satellite (TESS) \citep{Ricker+2015} and ground-based follow-up observations.
    We also search for helium in its atmosphere with high-resolution transmission spectroscopy. The paper is organized as follows. In Sections \ref{sec:TESS} and \ref{sec:obs}, we introduce the observation of TESS and ground-based telescope. In Sections \ref{sec:stellar}, \ref{sec:planet}, \ref{sec:MandR_results}, and \ref{sec:spectroscopy}, we describe the analysis and results for stellar parameters, planet parameters, and transmission spectroscopy. We discuss the physical properties in Section \ref{sec:dis}, and we summarize this work in Section \ref{sec:sum}.

\section{TESS photometry} \label{sec:TESS}

\begin {figure*} [htbp]
\begin{center}
 \includegraphics[width=\textwidth] {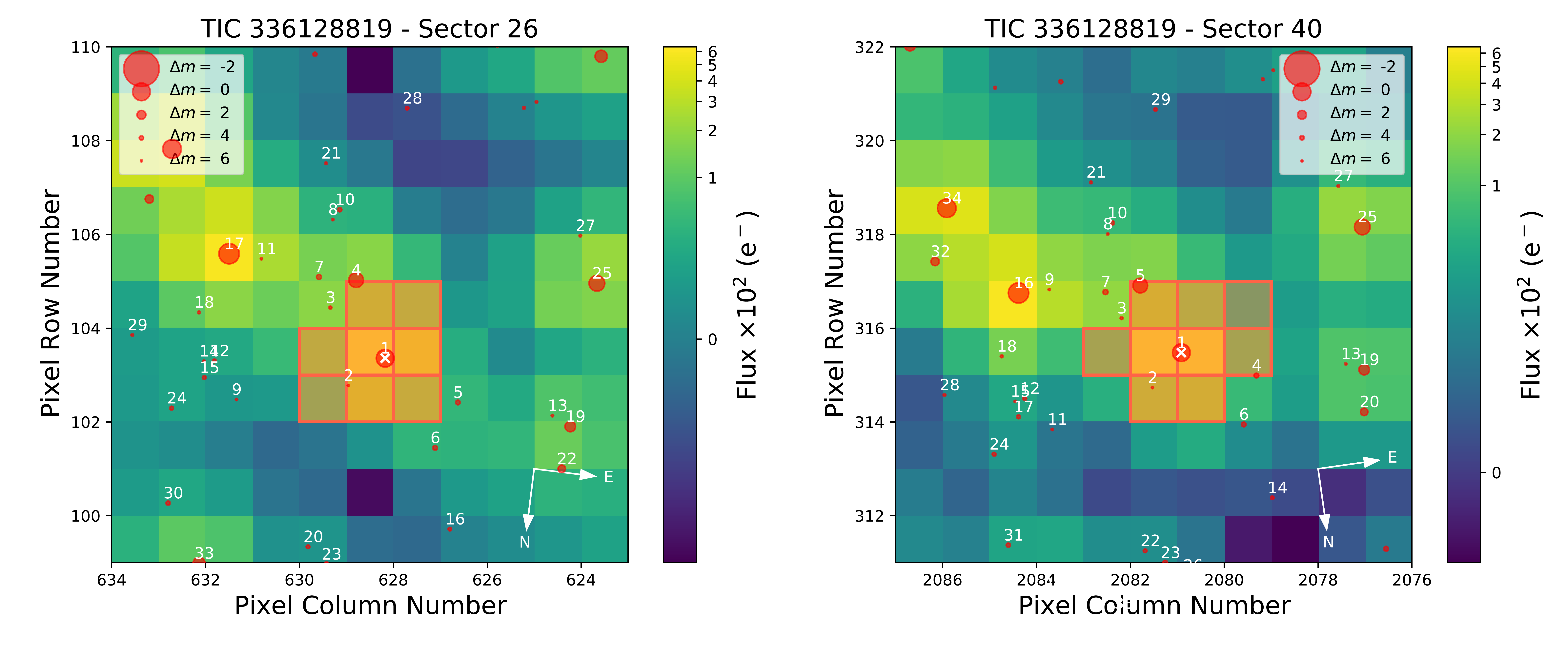}
\end{center}
\caption{TESS target pixel file image of TOI-2136 observed in Sector 26 (left) and Sector 40 (right). The pixels in red show the photometric aperture used by TESS to create the PDCSAP light curves. The position of nearby stars and their magnitudes are represented by the red circles. This image was produced using \texttt{tpfplotter} \citep{Aller+2020}.}
\label{fig:TPF_plot}
\end{figure*}

    TIC 336128819 (TOI-2136) was observed with a two-minute cadence by TESS in Sector 26 from UT 9 June 2020 to 4 July 2020, and in Sector 40 from UT 24 June 2021 to 23 July 2021. For each sector there are 0.919 days downtime to download data and 23.95 days of science data in total. For both TESS sectors, the target star was positioned on charge-coupled device (CCD) 1 Camera 1. Figure \ref{fig:TPF_plot} shows the TESS images around TOI-2136 for Sectors 26 and 40.

    TESS raw images were processed by the Science Processing Operations Center (SPOC) at NASA Ames Research Center. After the data were reduced by the SPOC pipeline \citep{Jenkins+2016}, a preliminary TESS light curve was produced using simple aperture photometry \citep[SAP;][]{Morris+2020} and instrumental systematic effects were removed using the presearch data conditioning (PDC) pipeline module \citep{Smith+2012,Stumpe+2014}. A transit search \citep{JenkinsJM2002, JenkinsJM2020} was carried out on the TESS light curves and a signal of $\sim 7.8$ days was detected. After the transit signal was validated by the TESS team \citep{Twicken2018}, it was assigned a TESS object of interest number (TOI-2136.01) and announced to the community. 

\section{Follow-up observations} \label{sec:obs}

\subsection{MuSCATs}

    We observed a total of five full transit events with  the Multicolor Simultaneous Camera for studying Atmospheres of Transiting exoplanets (MuSCAT) \citep{Narita+2015}, MuSCAT2 \citep{Narita+2019} and MuSCAT3 \citep{Narita+2020}. MuSCAT is mounted on the 1.88 m telescope of the National Astronomical Observatory of Japan (NAOJ) in Okayama, Japan. MuSCAT2 is mounted on the 1.52 m Telescopio Carlos S\'{a}nchez (TCS) at the Teide Observatory, in the Canary Islands, Spain. MuSCAT3 is mounted on the 2 m Faulkes Telescope North at the Haleakala Observatory, in Maui, Hawai'i. These instruments have $g$, $r$, $i$ and $\rm z_{s}$ bands (only $g$, $r$, and $\rm z_{s}$ bands for MuSCAT). MuSCAT and MuSCAT2 are equipped with 1024$\times$1024 pixel CCDs, and MuSCAT3 is equipped with 2048$\times$2048 pixel CCDs. MuSCAT provides a field of view (FOV) of 6.1 $\times$ 6.1 $\rm arcmin^2$ with a pixel scale of 0.36 arcsec per pixel. MuSCAT2 provides a field of view of 7.4 $\times$ 7.4 $\rm arcmin^2$ with a pixel scale of 0.44 arcsec per pixel. MuSCAT3 provides a field of view of 9.1 $\times$ 9.1 $\rm arcmin^2$ with a pixel scale of 0.27 arcsec per pixel.

    The first transit was observed by MuSCAT3 on UT 12 June 2021, and the exposure times were set to 60, 26, 23, and 20 s for $g$, $r$, $i$, and $\rm z_{s}$, respectively. The fourth transit was observed by MuSCAT on UT 23 September 2021, and the exposure times were set to 30 and 30 s for $r$ and $\rm z_{s}$; for that night the $g$-band CCD of MuSCAT did not work. 
The rest of the transits were observed by MuSCAT2 on UT 22 and 30 August 2021, and 24 October 2021. 
For the observations of 22 August 2021, the exposure times were set to 30, 15, and 15 s for $g$, $i$, and $\rm z_{s}$, respectively; for that night the $r$-band CCD of MuSCAT2 did not work and we also discarded the $g$-band data due to the large photometric scatter present in the light curve. 
For the MuSCAT2 observations of 30 August 2021, the exposure times were the same for each band as the night of 22 August 2021; the $r$-band CCD did not work that night also, but the $g$-band data were much better and we decided to include them in the analysis. 
For the night of 24 October 2021, the exposure times were set to 30, 20, 15, and 15 s for the $g$, $r$, $i$, and $\rm z_{s}$ bands, respectively. 
    We used a fixed aperture radius of 4.3 arcseconds for each band for the MuSCAT photometry, and 10.9 arcseconds for each band for the MuSCAT2 photometry. For the MuSCAT3 photometry, we used fixed aperture radii of 3.7, 4.3, 4.3, and 4.8 arcseconds for $g$, $r$, $i$, and $\rm z_{s}$, respectively.

\subsection{LCO 1 m / Sinistro}

    We observed one almost-full transit of TOI-2136b on UT 9 October 2021 simultaneously with two 1~m telescopes of Las Cumbres Observatory (LCO) at the McDonald Observatory in the USA. Each telescope is equipped with a single-band imager Sinistro, which has a 4k $\times$ 4k CCD with a pixel scale of 0\farcs389 pixel$^{-1}$ and an FOV of 26\farcm5 $\times$ 26\farcm5. We selected the $i$-band filter for one telescope and the $Z$-band filter for the other. We set the exposure time at 30~s with the full-frame readout mode for both telescopes. The telescopes were slightly defocused (by 0.2~mm) to avoid detector saturation. The obtained raw images were processed by the {\tt BANZAI} pipeline \citep{McCully2018} to perform dark-image and flat-field corrections. Aperture photometry was then performed using a custom pipeline \citep{2011PASJ...63..287F}. A summary of the ground-based photometric observations, facilities, instruments, and filters used in this work is presented in Table \ref{tab:groundobs}.

\begin{table}
\caption{Ground-based photometric observations.}             
\label{tab:groundobs}      
\centering  
\scalebox{0.9}{
\begin{tabular}{l c c c}        
\hline\hline                 
Date (UT) & Telescope/Instrument & Telescope size & Filters \\    
\hline                        
 2021-06-12 & Faulkes/MuSCAT3 & 2.0 m & $g$, $r$, $i$, $z_s$ \\
 2021-08-22 & TCS/MuSCAT2 & 1.5 m & $i$, $z_s$ \\
 2021-08-30 & TCS/MuSCAT2 & 1.5 m & $g$, $i$, $z_s$ \\
 2021-09-23 & Okayama/MuSCAT & 1.88 m & $r$, $z_s$ \\
 2021-10-08 & LCO/Sinistro & 1.0 m & $i$, $z$ \\
 2021-10-24 & TCS/MuSCAT2 & 1.5 m & $g$, $r$, $i$, $z_s$ \\
\hline                                   
\end{tabular}
}
\end{table}

\subsection{Subaru / IRD}

    We obtained a total of 38 high-resolution spectra of TOI-2136 from UT 27 September 2020 to 27 October 2021 with the InfraRed Doppler (IRD) instrument on the Subaru 8.2 m telescope under the Subaru-IRD TESS intensive follow-up program (ID: S20B-088I, S21B-118I). The IRD is a fiber-fed, near-infrared (NIR) spectrometer covering from 970 to 1750 nm with a spectral resolution of 70,000 \citep{Tamura+2012,Kotani+2018}. The exposure times were set to 750 - 1800 s depending on the sky condition. We also observed a telluric standard star (A0 or A1 star) to correct for telluric lines in the template spectrum, for the radial velocity (RV) analysis. 

    Raw IRD data were reduced with \texttt{IRAF} \citep{Tody1993} and our custom code to correct for the bias pattern of the detectors, and also to perform wavelength calibration with the Th-Ar lamp and laser-frequency comb \citep{Kuzuhara+2018,Hirano+2020}. The reduced one-dimensional spectra have signal-to-noise ratios (S/Ns) of 40 - 94 per pixel around 1000 nm. We measured the precision RV for each frame with the RV analysis pipeline for the IRD described in \citet{Hirano+2020}. The typical RV internal errors are 3 - 5 m $\rm s^{-1}$. The RV individual measurements are given in Table \ref{tab:RV_values} in the Appendix. 

    The IRD data also include the observation of a full transit of TOI-2136b on 27 September 2020, with the goal of searching for an excess helium absorption by the planetary atmosphere. We obtained 11 consecutive spectra with a 900 s exposure time from UT 05:01 to 07:51, out of which five frames are out-of-transit spectra and six frames are in-transit spectra. The S/Ns of the extracted spectra are 50 - 63 per pixel around 1000 nm, and the airmass varied from 1.05 to 1.33 during the observations. 

\subsection{Gemini North / ‘Alopeke}

    Exoplanet parameters, such as the radius determination, depend on the transit depth as well as the assumption that the star is single. A spatially close companion (bound or line-of-sight) can create a false-positive transit signal, or cause "third-light" flux, leading to an underestimation of the planet radius \citep{Ciardi+2015}. Such companion stars can even cause nondetections of small planets residing within the same exoplanetary system \citep{Lester+2021}. Nearly 50\% of FGK-type stars are binary or multiple star systems \citep{Matson+2018}, providing crucial information toward our understanding of exoplanetary formation, dynamics, and evolution \citep{Howell+2021}. Thus, to search for close-in ($<$1") companions unresolved in TESS images or other ground-based follow-up observations, we obtained high-resolution imaging speckle observations of TOI-2136.

\begin {figure} [htbp]
\begin{center}
 \includegraphics[width=0.5\textwidth] {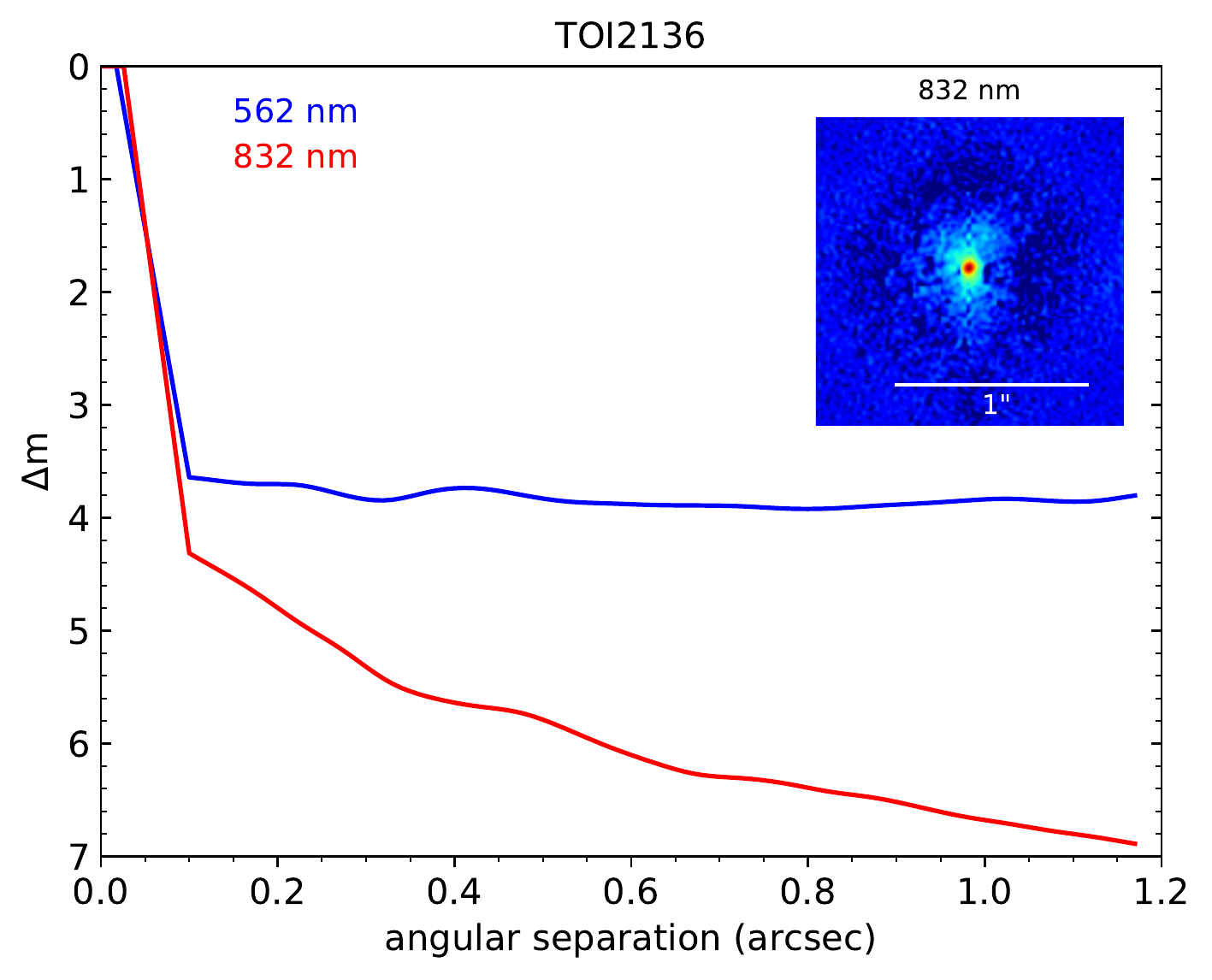}
\end{center}
\caption{Gemini/‘Alopeke 5-sigma contrast curves in 562 nm (blue) and 832 nm (red). The inset figure is the 832 nm reconstructed speckle image.}
\label{fig:Speckle}
\end{figure}

    TOI-2136 was observed on UT 17 October 2021 using the ‘Alopeke speckle instrument on the Gemini North 8-m telescope \citep{Scott+2021}. ‘Alopeke provides simultaneous speckle imaging in two bands (562 nm and 832 nm) with output data products including a reconstructed image with robust contrast limits on companion detections \citep[e.g.,][]{Howell+2016}. Five sets of 1000 $\times$ 0.06 sec exposures were collected for TOI-2136 and subjected to Fourier analysis in our standard reduction pipeline \citep[see][]{Howell+2011}. Figure \ref{fig:Speckle} shows our final contrast curves and the 832 nm reconstructed speckle image. We find that TOI-2136 is a single star with no companion dimmer than four to seven magnitudes above that of the target star from the diffraction limit (20 mas) out to 1.2”. At the distance of TOI-2136 (d=33 pc), these angular limits correspond to spatial limits of 0.7 to 40 au.

\section{Stellar properties} \label{sec:stellar}

\subsection{Stellar parameters}

    We determined the effective temperature $T_{\mathrm{eff}}$ and metallicity of the host star from the template spectrum, which was constructed from the IRD spectra. 
    The analysis is based on a line-by-line comparison of equivalent widths (EWs) between the observed spectra and synthetic spectra, following the procedure described in Section 3.1.2 of \citet{2021AJ....162..161H}.
    The synthetic spectra were calculated by a one-dimensional LTE spectral synthesis code, which was based on the same assumptions as the model atmosphere program of \citet{1978A&A....62...29T}, with the interpolated grid of MARCS model atmosphere \citep{2008A&A...486..951G}.
    Through spectral analysis, a surface gravity $\log{g}$ value of $4.88 \pm 0.0008$ was adopted from the TESS Input Catalog (TIC) version 8 \citep{Stassun+2019}, and the microturbulent velocity was fixed at $0.5 \pm 0.5$ km\,s$^{-1}$ for simplicity.

    The selected 47 FeH molecular lines in the Wing-Ford band at $990-1020$ nm were used for the $T_{\mathrm{eff}}$ estimation.
    \citet{2022AJ....163...72I} provides the selection criteria and the list of the 47 lines.
    The original data of these line transitions are available on the MARCS\footnote{\url{https://marcs.astro.uu.se/}} web page.
    We measured the EW of each FeH line by fitting the Gaussian profile to find the $T_{\mathrm{eff}}$ at which the synthetic spectra best reproduce the EW by an iterative search.

    For the stellar metallicity, we used the analysis tool developed by \citet{2020PASJ...72..102I}.
    We used the atomic absorption lines of Na, Mg, Ca, Ti, Cr, Mn, Fe, and Sr based on the selection criteria of: (1) not suffering from blending of other absorption lines, (2) sensitive to elemental abundances, and (3) continuum level can be reasonably determined.
    The original data of the atomic lines are taken from the Vienna Atomic Line Database \citep[VALD;][]{1999A&AS..138..119K,2015PhyS...90e4005R}.
    The EWs were measured by fitting the synthetic spectra on a line-by-line basis.
    We searched for an elemental abundance until the synthetic EW matched the observed one for each line, and took the average for all the lines to estimate [X/H] for an element X.
    The [M/H] is calculated as an error-weighted average of abundances of all the eight elements.

    As a first step, we estimated $T_{\mathrm{eff}}$ assuming the solar metallicity.
    The average of the $T_{\mathrm{eff}}$ estimates for the 47 lines was adopted as the best estimate.
    We temporarily determined the elemental abundances using the $T_{\mathrm{eff}}$ estimate and, as a second step, we redetermined $T_{\mathrm{eff}}$ adopting the iron abundance [Fe/H] as the metallicity of the atmospheric model grid.
    Its uncertainty was given as the sum of the statistical error (8 K), which was calculated by dividing the standard deviation of the 47 lines by the square root of the number of lines ($\sigma / \sqrt{N}$), and the possible systematic error (100 K).
    We then adopted the resulting $T_{\mathrm{eff}}$ to finally determined the elemental abundances.
    The procedure up to this point allowed the results of $T_{\mathrm{eff}}$ and abundances to converge well within the measurement errors.
    Consequently, we obtained $T_\mathrm{eff}=3373\,\pm\,108$ K, $\mathrm{[Fe/H]}=0.02\, \pm \,0.14$ dex, and $\mathrm{[M/H]}=0.06\, \pm \,0.06$ dex.

    Based on these parameters, as well as the literature values in Table~\ref{tab:sparam}, we estimated the other physical parameters, including the stellar radius, which is required to infer the planet radius. Making use of the empirical relations for M dwarfs by \citet{Mann+2015, Mann+2019}, we derived the stellar mass $M_*$, radius $R_*$, and density $\rho_*$ with their uncertainties estimated by the Monte Carlo approach using Gaussian distributions for the Gaia parallax, 2MASS magnitudes, and [Fe/H]. 
    The stellar luminosity $L_*$ was also derived from the Stefan-Boltzmann law using $T_\mathrm{eff}$ in Table~\ref{tab:sparam}.

\begin{table}
\caption{Stellar parameters.}             
\label{tab:sparam}      
\centering                          
\begin{tabular}{l c c}        
\hline\hline              
Parameter & Value & reference \\
\hline 
 TIC & 336128819 & \\
 $\alpha$ (J2000) & 18:44:42.37 & (a)\\
 $\delta$ (J2000) & 36:33:47.36 & (a)\\
 $\mu_{\alpha}\cos\delta$ (mas $\rm yr^{-1}$) & -33.809 $\pm$ 0.017 & (a)\\
 $\mu_{\delta}$ (mas $\rm yr^{-1}$) & 177.053 $\pm$ 0.020 & (a)\\
 parallax (mas) & 29.9756 $\pm$ 0.0169 & (a)\\
 Gaia & 12.9462 $\pm$ 0.0005 & (a)\\
 TESS & 11.7371 $\pm$ 0.0074 & (b) \\
 V mag & 14.32 $\pm$ 0.2 & (c)\\
 J mag & 10.184 $\pm$ 0.024 & (d)\\
 H mag & 9.604 $\pm$ 0.028 & (d)\\
 K mag & 9.343 $\pm$ 0.022 & (d)\\
 distance (pc) & 33.361 $\pm$ 0.019 & This work \\
 $T_\mathrm{eff}$ (K) & 3373 $\pm$ 108 & This work\\
 $U$ (km\,s$^{-1}$) & -35.20 $\pm$ 0.13 & This work\\
 $V$ (km\,s$^{-1}$) & -19.61 $\pm$ 0.29 & This work\\
 $W$ (km\,s$^{-1}$) & 6.05 $\pm$ 0.10 & This work\\
 $\left[\rm Fe/H\right]$ (dex) & 0.02 $\pm$ 0.14 & This work\\
 $\left[\rm M/H\right]$ (dex) & 0.06 $\pm$ 0.06  & This work\\
 $\log g$ (cgs) & 4.881 $\pm$ 0.026 & This work\\
 $M_{*}$ (M$_\odot$)& 0.3272 $\pm$ 0.0082 & This work\\
 $R_{*}$ (R$_\odot$) & 0.3440 $\pm$ 0.0099 & This work\\
 $\rho_{*}$ ($\rho_\odot$) & 8.04$^{+0.74}_{-0.66}$ & This work\\
 $L_{*}$ (L$_\odot$) & 0.0137$^{+0.0019}_{-0.0017}$ & This work\\
 $P_{rot}$ (days) & $82.56 \pm 0.45$ & This work\\
\hline                                   
\end{tabular}
\tablefoot{(a) GAIA EDR3 \citep{GAIADR3+2021}, \ (b) \citet{Stassun+2019}, (c) \citet{Lepine2005}, (d) \citet{Cutri+2003}.}
\end{table}

\subsection{Stellar rotation} \label{sec:StellarRot}

    We obtained publicly available photometry of TOI-2136 from the MEarth project (\citealp{Berta2012}) and the Zwicky Transient Facility (ZTF, \citealp{Masci2019}). The MEarth project uses eight 40 cm telescopes equipped with RG715 filters and it mostly focuses on observing M stars to search for transits (\citealp{Nutzman2008}). The project has two observing sites in both hemispheres. One site is located at the Fred Lawrence Whipple Observatory, on Mount Hopkins in Arizona, in the USA. The other is located at the Cerro Tololo Inter-american Observatory (CTIO) in Chile. From the MEarth project, we obtained TOI-2136 (LSPMJ1844+3633) photometry from the 2011-2020 north target light curves. 

    The ZTF uses a 47 squared degree camera mounted on the Palomar 48-inch Schmidt Telescope to search and study transient and variable objects in the sky. Using its large observing area, ZTF is capable of scanning the northern hemisphere sky every two nights. From ZTF we obtained $g-$ and $r-$band photometry of TOI-2136 from the ZTF data release 8 (DR8). 

\begin {figure*} [htbp]
\begin{center}
 \includegraphics[width=\textwidth] {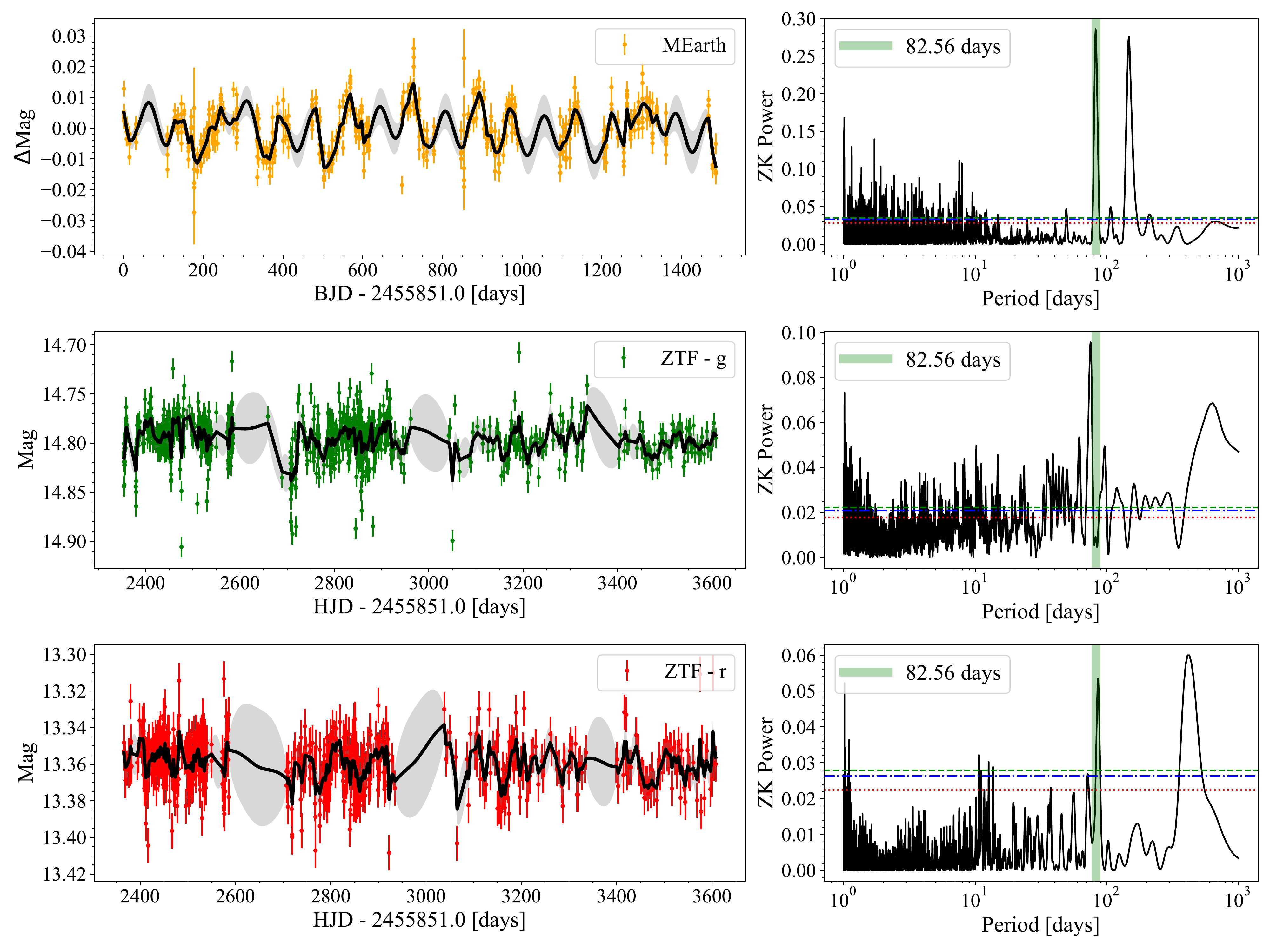}
\end{center}
\caption{Ground-based long-term photometric observations of TOI-2136 using MEarth and ZTF, and their respective GLS periodograms (\citealp{Zechmeister2009}) for each data set. The stellar photometric variability is detected with a stronger signal in MEarth data. Fitting a kernel with a periodic term to all the data sets using Gaussian processes, we find a stellar rotation period of $82.56 \pm 0.45$ days.}
\label{fig:RotationalPeriod}
\end{figure*}

    Visual inspection of the MEarth and ZTF data shows that TOI-2136 presented periodic photometric variations (see Fig. \ref{fig:RotationalPeriod}). These variations can be produced by inhomogeneities in the stellar surface (e.g., spots, plages) that appear and disappear out of sight as the star rotates. We performed a joint fit to the MEarth and ZTF photometry to establish the rotation period of the star. We used the package for Gaussian Processes \texttt{Celerite} (\citealp{ForemanMackey2017}) and fit the photometry with the following kernel:
\begin{equation}
    k_{ij\; \mathrm{Phot}} = \frac{B}{2+C} e^{-|t_i-t_j|/L} \left[  \cos \left( \frac{2 \pi |t_i-t_j|}{P_{rot}} \right) + (1 + C)  \right]
\label{Eq:StarRot_GPKernel}
,\end{equation}
    where $|t_i-t_j|$ is the difference between two epochs or observations, $B$, $C$, and $L$ are positive constants, and $P_{rot}$ is the stellar rotational period (see \citealp{ForemanMackey2017} for details). For the joint fit, each data set had the constants $B$, $C$, and $L$ as free parameters, but shared a common rotational period. The prior functions and parameter limits used in the joint fit are in Table \ref{tab:Prot_priors} in the Appendix. After global optimization of a log likelihood function, we sampled the posterior distribution of the kernel used to fit the photometry with \texttt{emcee} (\citealp{ForemanMackey2013}), using 80 chains and 20000 iterations. The final parameter values (median and 1$\sigma$ uncertainties) were estimated from the posterior distribution.

    Figure \ref{fig:RotationalPeriod} shows the MEarth and ZTF photometry, and the best fitted model using the kernel described in Eq. \ref{Eq:StarRot_GPKernel}. The generalized Lomb-Scargle (GLS, \citealp{Zechmeister2009}) periodograms for each photometric data set are shown in the right panel. The three data sets present peaks in their respective periodograms for periods between 70 and 90 days. The stellar rotational period found by the best fitted model is $P_{rot} = 82.56 \pm 0.45$ days. This rotation period is in agreement with \cite{Newton2016}, which found a rotational period of $P_{rot} = 82.97$ days for this star, using MEarth data.

    \cite{Engle+2018} investigated the relationship between the stellar age and rotation of M dwarfs. Using their equation, the stellar age of TOI-2136 is calculated as $\sim $ 5 Gyr. We also estimated the stellar age from its correlation with the Galactic space velocities ($U$, $V$, $W$). We computed the $UVW$ velocities using the Gaia Early Data Release (EDR) 3 information and the absolute RV measured from the IRD spectra using the procedure described in \cite{Johnson+1987} (Table~\ref{tab:sparam}). Following \cite{2021AJ....162..161H}, we obtained an age of $\rm 6.2^{+3.7}_{-3.1}$ Gyrs when the age prior based on the Geneva-Copenhagen Survey catalog \citep{Casagrande+2011} is imposed, and $\rm 6.3^{+5.0}_{-4.1}$ Gyrs for the uniform age prior (0 < age < 14 Gyr). These results indicate that the stellar rotation and $UVW$ velocities do not constrain the age of the star.

\section{Methods} \label{sec:planet}

\subsection{Transit light curve modeling} \label{sec:tarnsit}

    For the transit and RV modeling, we followed the procedure described in \cite{Murgas2021}. The light curves were modeled using the \texttt{python} package \texttt{PyTransit}\footnote{\url{https://github.com/hpparvi/PyTransit}} (\citealp{Parviainen2015}). To fit the transit events analyzed here, we used \texttt{PyTransit's} \cite{Mandel2002} analytic models, assuming a quadratic limb-darkening law.

    Before performing the joint fit of all the data available, the ground-based transit observations were fitted simultaneously using a common transit model and for each night the systematic effects were accounted for independently. To model the instrumental noise, we used a linear model with one free parameter dependent on the airmass of the target, two free parameters dependent on the star position in the detector (X- and Y-axis), and a term dependent on the full width at half maximum (FWHM) of the point spread function (PSF) as a proxy for seeing variations. For the final joint fit we used the LCO, MuSCAT, MuSCAT2 and MuSCAT3 detrended light curves (i.e., without systematic effects).

    For the joint fit, the common transit free parameters used in the transit modeling were the planet-to-star radius ratio $\mathrm{R}_\mathrm{p}/\mathrm{R}_*$, the central time of the transit $\mathrm{T}_{\mathrm{c}}$, the stellar density $\rho_*$, and the transit impact parameter $\mathrm{b}$. A constant orbital period (i.e., assuming no transit timing variations) was used as a common parameter for the transit light curves and RV measurements. The passband dependent quadratic limb-darkening (LD) coefficients $\mathrm{u}_1$ and $\mathrm{u}_2$ were set free. During the fitting process, we converted the LD coefficients $(\mathrm{u}_1, \mathrm{u}_2)$ to the parametrization proposed by \cite{Kipping2013}, $(\mathrm{q}_1,\mathrm{q}_2)$ with $q_i \in [0,1]$. The fitted LD coefficients were weighted against the predicted coefficients delivered by the \texttt{Python Limb Darkening Toolkit}\footnote{\url{https://github.com/hpparvi/ldtk}} (\texttt{ldtk}, \citealp{Parvianine2015LDTK}), computed using the TOI-2136 stellar parameters from Table \ref{tab:sparam}.

    To model the stellar variability and residual systematic noise present in the TESS data, we used Gaussian processes (GPs; e.g., \citealp{Rasmussen2006}, \citealp{Gibson2012}, \citealp{Ambikasaran2015}). For the GPs used in the TESS data, we chose the Matern $3/2$ kernel implementation of \texttt{Celerite} (\citealp{ForemanMackey2017}):
\begin{equation}
    k_{ij\; \mathrm{TESS}} = c^2_1 \left( 1 + \frac{\sqrt{3} |t_i-t_j|}{\tau_1}\right) \exp\left(-\frac{\sqrt{3} |t_i-t_j|}{\tau_1}\right)
\label{Eq:TESS_GPKernel}
,\end{equation}
    where $|t_i-t_j|$ is the difference between two epochs or observations, $c_1$ is the amplitude of the flux variation, and $\tau_1$ is a characteristic timescale. For each TESS data set available, the constants $c_1$ and $\tau_1$ were set as free parameters in the fit.

\subsection{Radial velocity modeling} 
\label{sec:spectrum}

    The RV data were modeled using the Radial Velocity Modeling Toolkit \texttt{RadVel}\footnote{\url{https://github.com/California-Planet-Search/radvel}} (\citealp{Fulton2018}). The free parameters used to model the RVs were: the planet-induced RV semiamplitude ($K_\mathrm{RV}$), the host star systemic velocity ($\gamma_0$), the orbital eccentricity and argument of the periastron (using the parametrization of $\sqrt{e}\cos \omega$, $\sqrt{e}\sin \omega$), and the instrumental RV jitter ($\sigma_{\mathrm{RV\; jitter}}$). The orbital period and central transit time were also set free, but were taken to be global parameters in common with the light curves. To account for systematic noise present in the RV time series, we used GPs with an exponential squared kernel (i.e., a Gaussian kernel):
\begin{equation}
    k_{ij\; \mathrm{RV}} = c^2_2 \exp \left(  - \frac{ (t_i-t_j)^2}{\tau^2_2} \right)
\label{Eq:RV_GPKernel}
,\end{equation}
    where $t_i-t_j$ is the difference between two epochs or observations, $c_2$ is the amplitude of the exponential squared kernel, and $\tau_2$ is a characteristic timescale. The constants $c_2$ and $\tau_2$ were set as free parameters.

    We tested how the fitted parameter values changed by fitting a single-planet keplerian to the data without taking into account the red noise in the RV measurements, and by modeling the systematics using GPs. We then computed the model comparison metrics Bayesian information criterion (BIC, \citealp{Schwarz1978}) and Akaike information criterion (AIC, \citealp{Akaike1974}), defined by
    \begin{equation}
    \mathrm{BIC} = k \ln{n} - 2\ln{\mathcal{L}_\mathrm{max}}\;.
    \label{Eq:BIC}
    \end{equation}
    and
    \begin{equation}
    \mathrm{AIC} = 2k - 2\ln{\mathcal{L}_\mathrm{max}}\;.
    \label{Eq:AIC}
    \end{equation}
    were $n$ is the number of observed data points, $k$ is the number of fitted parameters, and $\mathcal{L}_\mathrm{max}$ is the maximized likelihood. We found that the RV model that includes GPs is slightly preferred, although the difference between the criterion values for the RV model including GPs and without GPs ($\Delta \mathrm{BIC}$ and $\Delta \mathrm{AIC}$) is not large enough to select the single keplerian model plus the GPs model according to the criteria of \cite{Raferty1995} ($\Delta \mathrm{BIC} < 2.0$ and $\Delta \mathrm{AIC} < 2.5$).

\subsection{Joint fit} \label{sec:joint}

    For the joint fit of all the data available, we employed a Bayesian approach. We performed an uninformative transit search in the TESS time series (from Sectors 26 and 40) using \texttt{Transit Least Squares} (\texttt{TLS}, \citealp{Hippke2019}). From \texttt{TLS} we obtained an estimate of the period and epoch of the central time of the transit with their respective uncertainties; we used these values as priors for the joint fit. Then, we implemented a Markov chain Monte Carlo (MCMC) to explore the posterior distribution of the parameters using \texttt{emcee} (\citealp{ForemanMackey2013}). With \texttt{emcee} we evaluated a likelihood plus a prior function for several iterations. The likelihood function was the sum of the log likelihood for each transit time series plus a log likelihood for the star LD coefficients and the log likelihood of the RV observations. A total of 26 free parameters were sampled in the joint fit. The fitting process started with the optimization of the global likelihood with \texttt{PyDE}\footnote{\url{https://github.com/hpparvi/PyDE}}, and once the optimization process ended we used the optimal parameter distribution to start the MCMC. The MCMC consisted of 200 chains and ran for 2000 iterations as a burn-in, and the main MCMC ran for 8000 iterations. 

      The final parameter values and their respective uncertainties were computed from the posterior distributions, adopting the median as parameter values and using the 1$\sigma$ levels of the distribution as uncertainties. The prior functions and parameter limits used in the joint fit are in Table \ref{tab:pparam_priors} in the Appendix. Figure \ref{Fig:Fit_ParamDistr_CornerPlot} shows the correlation plots of the fitted orbital parameters, excluding the limb-darkening coefficients and parameters related to the modeling of red noise present in the data.

\section{Results} 
\label{sec:MandR_results}

    We present the fitted and derived parameters for the joint fit of the data with and without the use of GPs to model the systematics present in the RV measurements (Table \ref{tab:pparam}).
    For the case of the fit without GPs to model the red noise, the RV jitter parameter increased by a factor of two when compared to the GP modeling results; an expected result since there was no component to model the red noise in this approach.
    For the non-GP approach, we find an upper limit for the planetary mass of 8.3 $\rm M_\oplus$ while for the fit including GPs in the RV modeling, we find $\rm M_p < 9.9 $ $\rm M_\oplus$. As discussed in Section 5.2, the fits performed with and without using GPs for the RV measurements are not statistically different according to the difference between the model selection metrics estimated using the BIC and AIC definitions. We choose to adopt the results of the fit using GPs, since this approach takes into account the red noise in the data and the results provide a more conservative upper mass limit for the mass of the planet.

\begin {figure*} [htbp]
\begin{center}
 \includegraphics[width=\textwidth] {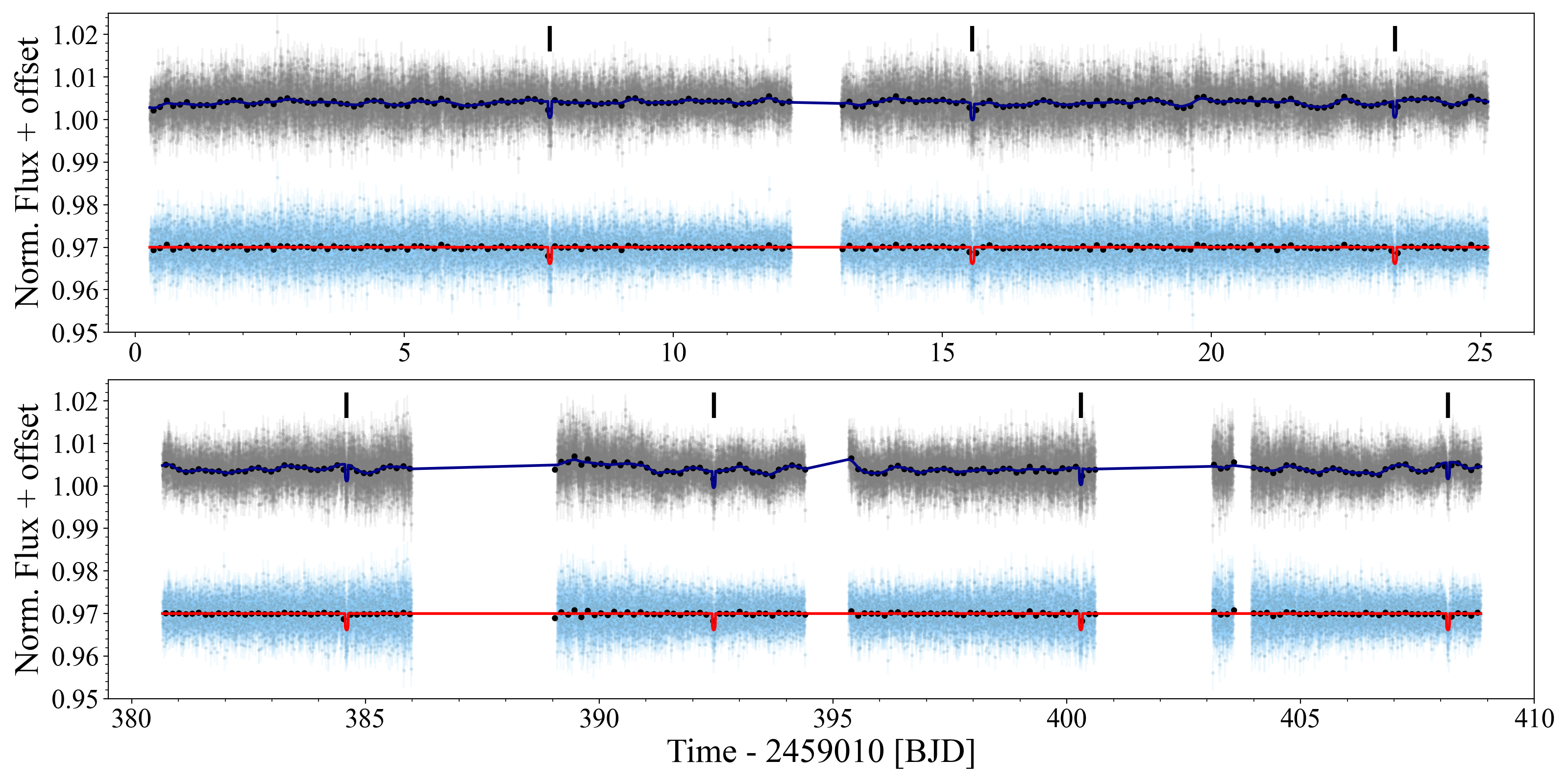}
\end{center}
\caption{TOI-2136 light curves of TESS Sector 26 (top panel) and Sector 40 (bottom panel). The black points represent TESS binned photometry. The gray points are the TESS PDCSAP data. The best fitting transit model model including systematic effects is shown in dark blue. Light-blue points are the TESS data after correcting the systematic effects. The best fitting transit model is shown in red. Transit events of TOI-2136b are marked with black vertical lines.}
\label{fig:TESS_LC_plot1}
\end{figure*}

\begin {figure*} [htbp]
\begin{center}
 \includegraphics[width=\textwidth] {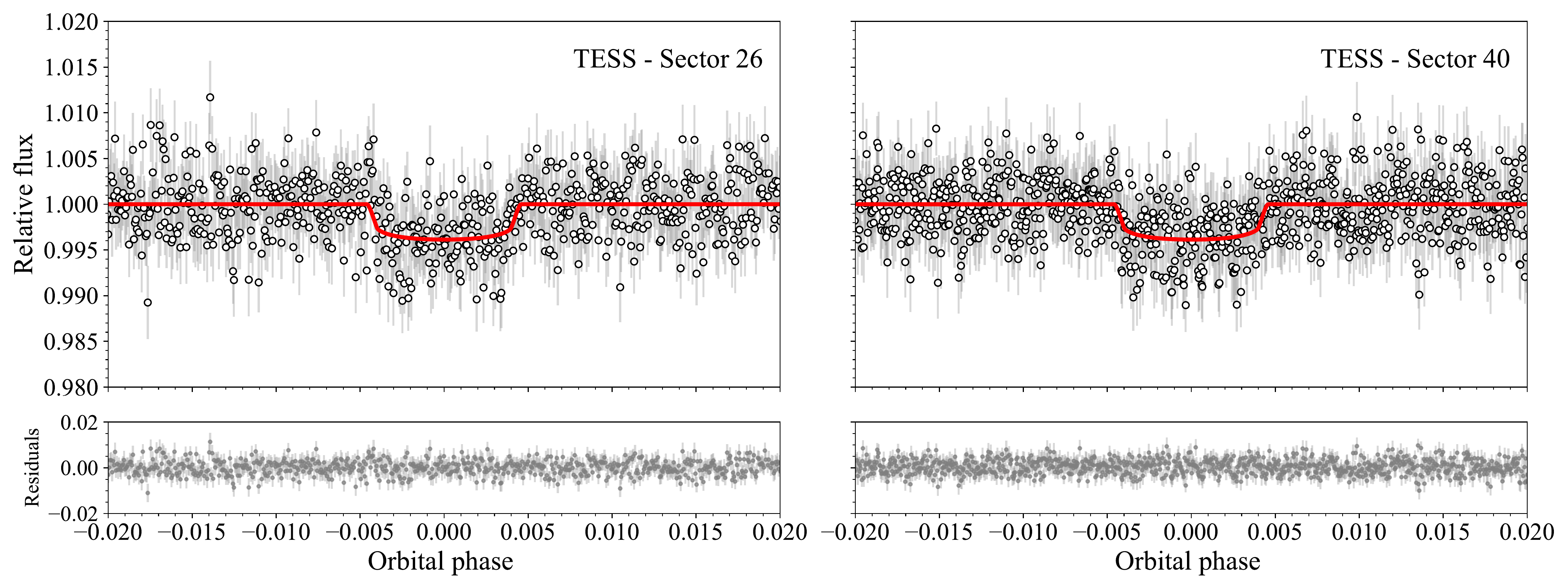}
\end{center}
\caption{TESS sector 26 and 40 folded light curves, and the best fitting model. The photometric variability present in both TESS sectors have been removed.}
\label{fig:TESS_LC_plot2}
\end{figure*}

\begin {figure*} [htbp]
\begin{center}
 \includegraphics[width=\textwidth] {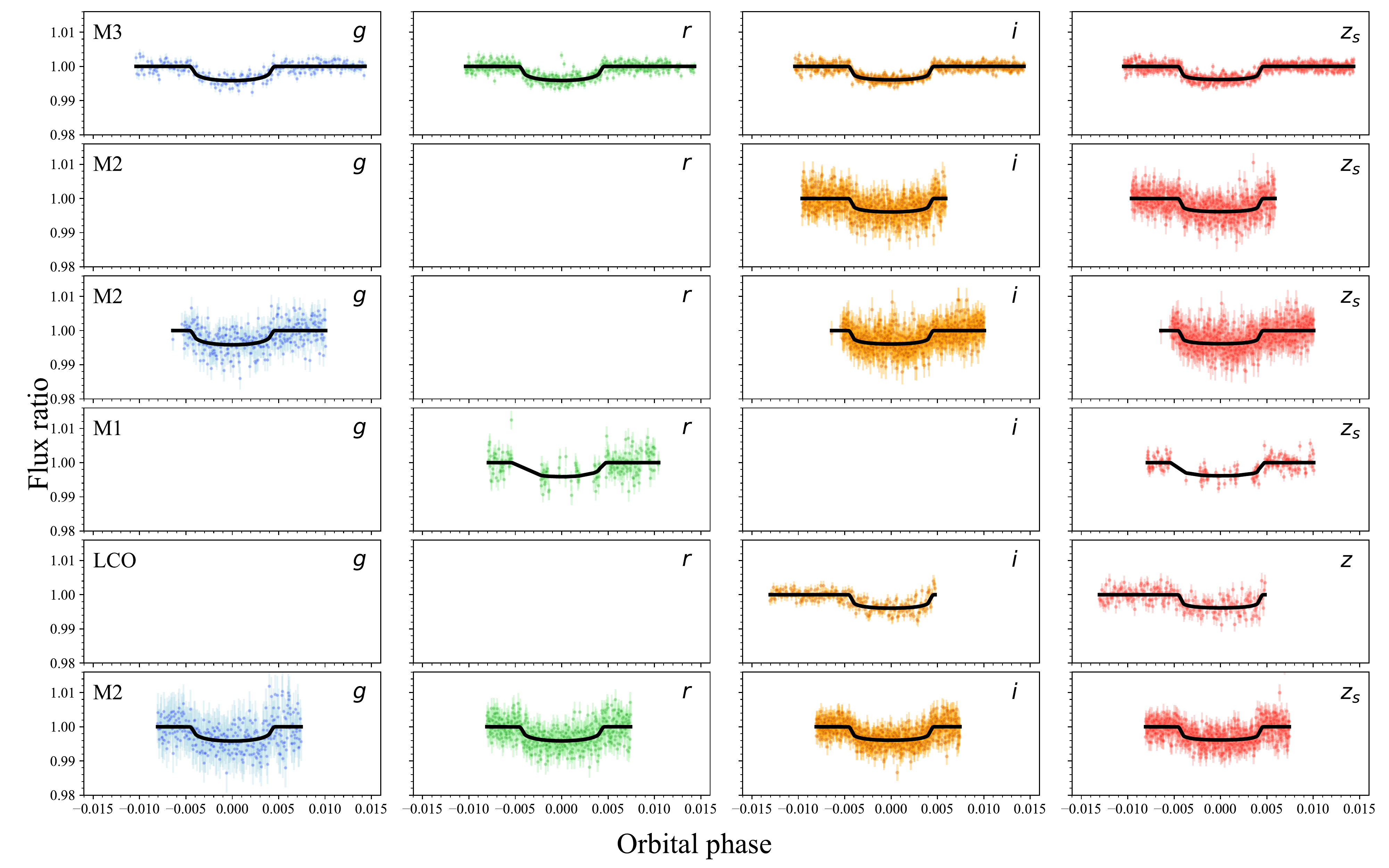}
\end{center}
\caption{Ground-based transit observations of TOI-2136b. The figure shows the photometry obtained with MuSCAT (M1), MuSCAT2 (M2), MuSCAT3 (M3), and LCO facilities, and the best fitting model (black line). The systematic effects presented in each data set were fitted and subtracted before performing the joint fit. The dates for each observation are in Table \ref{tab:groundobs}. The oldest observations are in the top row, and the most recent are in bottom row.}
\label{fig:MuSCAT_LC_plot}
\end{figure*}

\begin {figure*} [htbp]
\begin{center}
 \includegraphics[width=\textwidth] {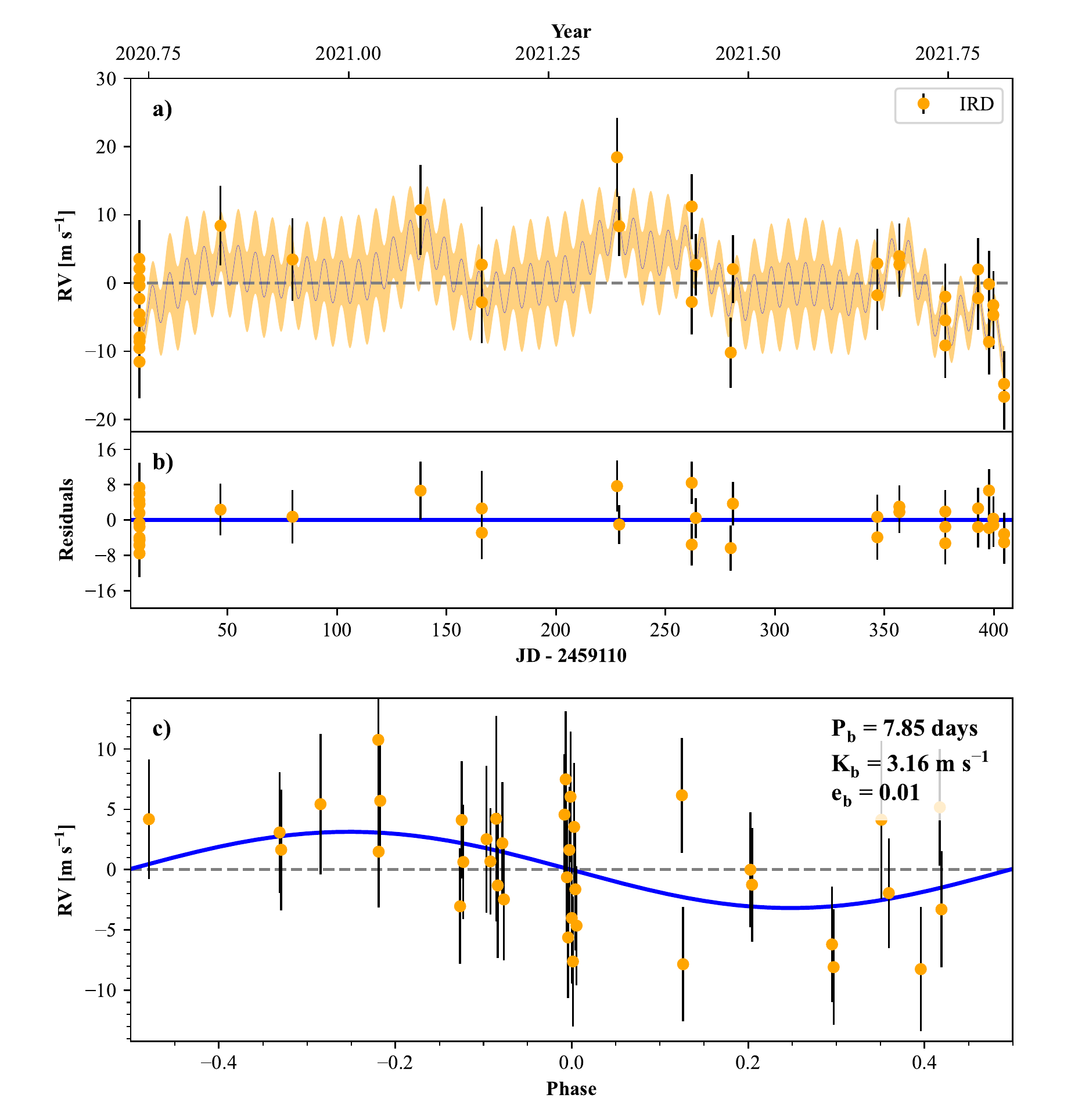}
\end{center}
\caption{Radial velocity measurements of TOI-2136 taken with the IRD. The top panel presents the time series and best fitting model with the use of GPs to model the red noise (the fit without GPs is presented in Fig. \ref{fig:Subaru_RV_plot_withoutGPs}). The middle panel presents the residuals of the fit. The bottom panel shows the RV measurements in phase.}
\label{fig:Subaru_RV_plot}
\end{figure*}

    Figure \ref{fig:TESS_LC_plot1} presents the photometric light curves from TESS Sectors 26 and 40, and the best fitting model (including systematic effects) found by the joint fit. Figure \ref{fig:TESS_LC_plot2} shows the detrended and phase-folded TESS light curves for both TESS sectors analyzed in this work. Figure \ref{fig:MuSCAT_LC_plot} shows the multicolor ground-based transit follow-up observations of TOI-2136b used in the joint fit. Finally, Figure \ref{fig:Subaru_RV_plot} presents the IRD RV measurements and the best fitting model. 

\begin{table*}
\caption{Planet parameters.}             
\label{tab:pparam}      
\centering                          
\begin{tabular}{l c c}        
\hline\hline   
Parameter & Joint fit - no GPs for RVs & Joint fit with GPs  \\
\hline
\multicolumn{3}{c}{Fitted orbital and transit parameters} \\
$R_{p}/R_{*}$ & $0.0588^{+0.0006}_{-0.0006}$ & $0.0588^{+0.0006}_{-0.0006}$ \\
$T_{c}$  [BJD] & $2459214.00325^{+0.00043}_{-0.00040}$ & $2459214.00322^{+0.00045}_{-0.00042}$ \\
$P$ [days] & $7.851925^{+0.000015}_{-0.000016}$ & $7.851925^{+0.000016}_{-0.000016}$ \\
$\rho_*$ [g cm$^{-3}$] & $11.33^{+0.69}_{-0.67}$ & $11.38^{+0.70}_{-0.71}$ \\
$b$ & $0.463^{+0.063}_{-0.072}$ & $0.462^{+0.067}_{-0.078}$ \\
$\sqrt{e}\cos(\omega)$ & $0.070^{+0.284}_{-0.270}$ & $-0.086^{+0.290}_{-0.251}$ \\
$\sqrt{e}\sin(\omega)$ & $-0.024^{+0.155}_{-0.143}$ & $-0.029^{+0.160}_{-0.145}$ \\
$\gamma_0 - \langle \gamma_0 \rangle$ [m/s] & $-1.89^{+1.23}_{-1.22}$ & $0.10^{+2.53}_{-2.49}$ \\
$K$ [m/s] & $2.53^{+1.92}_{-1.48}$ & $3.29^{+2.14}_{-1.80}$ \\
$\sigma_{RV}$ [m/s] & $5.89^{+1.18}_{-1.08}$ & $2.49^{+1.63}_{-1.56}$ \\
\multicolumn{3}{c}{Derived orbital parameters} \\
$e$ & $0.07^{+0.11}_{-0.05}$ & $0.07^{+0.09}_{-0.05}$ \\
$\omega$ [deg] & $-8.32^{+120.09}_{-113.43}$ & $-18.24^{+160.32}_{-140.29}$ \\
$a/R_*$ & $33.30^{+0.66}_{-0.67}$ & $33.35^{+0.67}_{-0.71}$ \\
$i$ [deg] & $89.20^{+0.11}_{-0.09}$ & $89.20^{+0.12}_{-0.09}$ \\
\multicolumn{3}{c}{Fitted LD coefficients} \\
$u_{1\;TESS}$ & $0.29^{+0.02}_{-0.02}$ & $0.28^{+0.02}_{-0.02}$ \\
$u_{2\;TESS}$ & $0.26^{+0.03}_{-0.04}$ & $0.26^{+0.04}_{-0.04}$ \\
$u_{1\;g}$ & $0.52^{+0.04}_{-0.04}$ & $0.52^{+0.03}_{-0.03}$ \\
$u_{2\;g}$ & $0.30^{+0.06}_{-0.06}$ & $0.29^{+0.05}_{-0.05}$ \\
$u_{1\;r}$ & $0.54^{+0.03}_{-0.03}$ & $0.54^{+0.03}_{-0.03}$ \\
$u_{2\;r}$ & $0.22^{+0.06}_{-0.05}$ & $0.22^{+0.06}_{-0.05}$ \\
$u_{1\;i}$ & $0.34^{+0.02}_{-0.01}$ & $0.34^{+0.02}_{-0.02}$ \\
$u_{2\;i}$ & $0.29^{+0.03}_{-0.03}$ & $0.29^{+0.04}_{-0.04}$ \\
$u_{1\;z_s}$ & $0.25^{+0.01}_{-0.01}$ & $0.24^{+0.02}_{-0.02}$ \\
$u_{2\;z_s}$ & $0.29^{+0.03}_{-0.03}$ & $0.30^{+0.03}_{-0.03}$ \\
\multicolumn{3}{c}{Fitted GP parameters} \\
$\log(c_1)$ TESS S26 & $-7.56^{+0.10}_{-0.10}$ & $-7.54^{+0.11}_{-0.10}$ \\
$\log(\tau_1)$ TESS S26 & $-1.61^{+0.18}_{-0.18}$ & $-1.60^{+0.18}_{-0.18}$ \\
$\log(c_1)$ TESS S40 & $-7.13^{+0.12}_{-0.11}$ & $-7.13^{+0.12}_{-0.11}$ \\
$\log(\tau_1)$ TESS S40 & $-1.21^{+0.16}_{-0.16}$ & $-1.21^{+0.16}_{-0.16}$ \\
$c_2$ IRD & ---- & $6.78^{+2.63}_{-1.88}$ \\
$\tau_2$ IRD & ---- & $13.72^{+21.52}_{-6.69}$ \\
\multicolumn{3}{c}{Derived planet parameters} \\
$R_{p}$ [$\rm R_{\oplus}$] & $2.21 \pm 0.07$ & $2.20 \pm 0.07$ \\
$M_{p}$ [$\rm M_{\oplus}$] & $3.7^{+2.8}_{-2.2}$ & $4.7^{+3.1}_{-2.6}$ \\
                       & $<8.3$ (95\%) & $<9.9$ (95\%) \\ 
$\rho_{p}$ [g cm$^{-3}$] & $1.91^{+1.46}_{-1.13}$ & $2.39^{+1.61}_{-1.36}$ \\
$g_p$ [m s$^{-2}$] & $7.5^{+5.7}_{-4.4}$ & $9.4^{+6.3}_{-5.3}$ \\
$a$ [au] & $0.0533 \pm 0.0015$ & $0.0533 \pm 0.0015$ \\
$T_{eq}$ [K] & $378 \pm 13$ & $378 \pm 13$ \\
$\langle F_{p} \rangle$ [10$^5$ W/m$^2$] & $0.066 \pm 0.010$ & $0.066 \pm 0.010$ \\
$S_{p}$ [$S_\oplus$] & $4.83 \pm 0.73$ & $4.82 \pm 0.72$ \\
\hline                                   
\end{tabular}
\tablefoot{$\mathrm{T}_\mathrm{eq}$ computed assuming a Bond albedo of 0.3.}
\end{table*}

    Our joint fit finds that TOI-2136b has a period of $P = 7.851925 \pm 0.000016$ days and an orbital eccentricity consistent with 0. We also determined that TOI-2136b has a radius of $R_{p} = 2.20 \pm 0.07$ $\rm R_{\oplus}$ and an equilibrium temperature of $T_{eq} = 378 \pm 13$ K.
    We note TOI-2136 light curves of TESS Sector 26 have a nontrivial bias by sky background noise correction. As this bias is about 1.54 \% in the data of Sector 26, it is only about 0.19 \% in our results, which is below the size of our error bars.

\begin {figure} [htbp]
\begin{center}
 \includegraphics[width=0.5\textwidth] {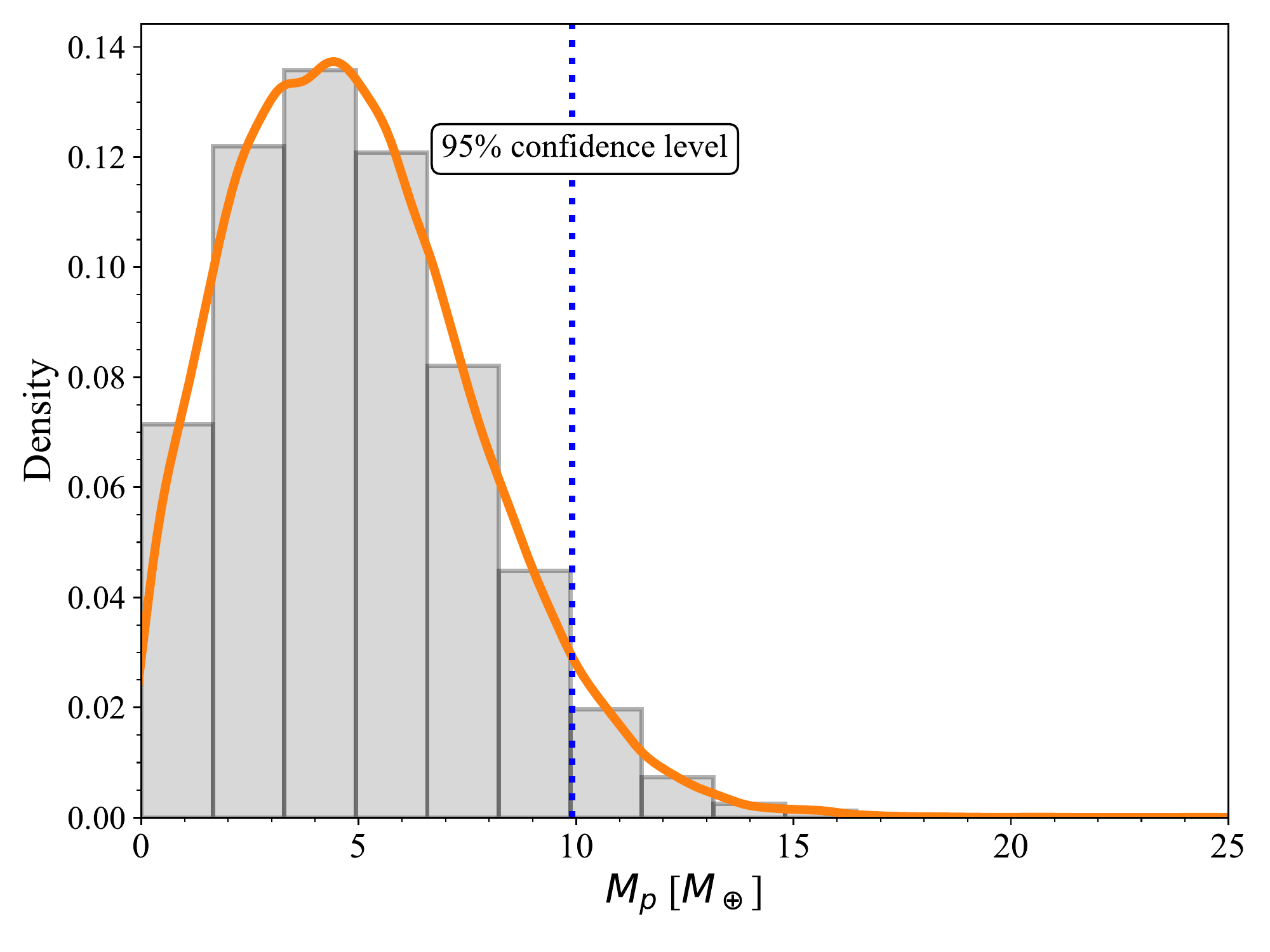}
\end{center}
\caption{Planetary mass posterior distribution. We can place an upper limit on the mass of TOI-2136b of 9.9 $\rm M_\oplus$ (95\% confidence level, dashed blue line).}
\label{fig:mass_postdistri}
\end{figure}

    From Figure \ref{fig:Subaru_RV_plot} it is clear that the RV measurements present some variability, the origin of which it is not clear. Looking at the long-term photometric follow-up (see Section \ref{sec:StellarRot}) and the TESS light curves, the star seems to present a low level of activity in the form of flares that could affect the precision of the RV measurements. To be conservative, we decided to fit a simple single-planet model to the RV data and assume that the rest of the signal is attributable to systematic noise modeled by the GPs. Figure \ref{fig:mass_postdistri} presents the posterior distribution for the planetary mass. The upper-limit (95\% confidence level) for the mass of TOI-2136b is 9.9 $\rm M_{\oplus}$. The planetary mass estimate from the joint fit is of $M_{p} = 4.7^{+3.1}_{-2.6}$ $\rm M_{\oplus}$. 
  
\begin {figure*} [htbp]
\begin{center}
 \includegraphics[width=\textwidth] {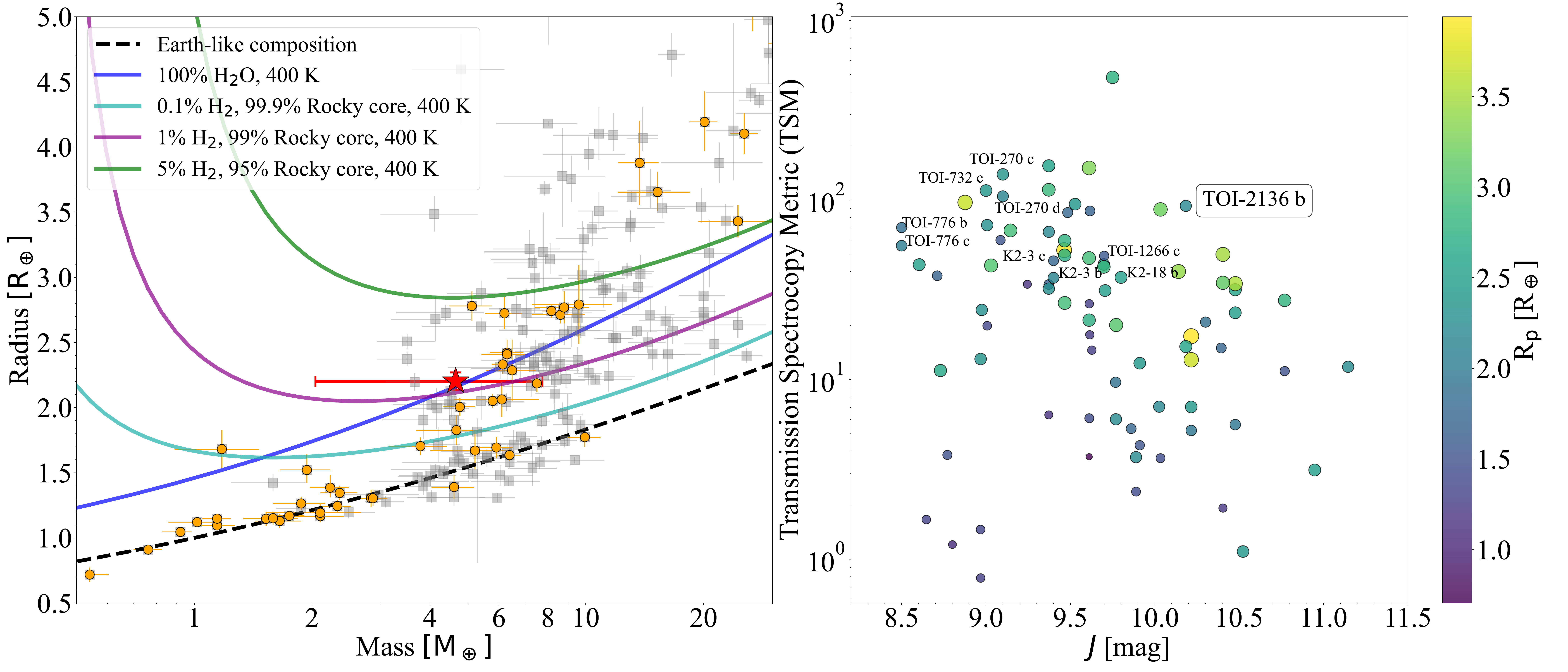}
\end{center}
\caption{Mass-radius diagram and TSM for TOI-2136b. Left panel: Mass and radius values for known exoplanets (data taken from TEPcat database, \citealp{Southworth2011}). The position of TOI-2136b is marked by the red star. In the figure we only show planets having mass measurements with a precision better than 30\%. 
The lines represent theoretical mass-radius relationships for five different bulk compositions, including bare rocky planets with an Earth-like composition, pure water worlds, and Earth-like rocky cores with 0.1\%, 1\%, and 5\% H+He envelopes with radiative equilibrium temperature of 400~K (see text).
The orange points represent planets orbiting M-type stars, and the gray squares are planets around other types of stars. 
Right panel: Apparent magnitude in \textit{J}-band vs the TSM by \cite{Kempton2018} for TOI-2136b and known transiting planets with radii smaller than 4 $\rm R_\oplus$. The other labeled planets are the hycean candidates.}
\label{fig:MassRadius_TSM}
\end{figure*}

    The left panel in Figure \ref{fig:MassRadius_TSM} shows the median mass and radius values of TOI-2136b derived in this work compared to known transiting exoplanets with precisely determined bulk properties. The parameters for the known exoplanets were taken from TEPcat database \citep{Southworth2011}. We only show planets having mass measurements with a precision better than 30\%. 
    In the panel we show theoretical mass-radius relationships for planets with five different bulk compositions, including bare rocky planets with an Earth-like composition (32.5\% Fe plus 67.5\% MgSiO$_3$), 100\% water worlds, and Earth-like rocky cores with 0.1\%, 1\%, and 5\% H+He envelopes. For bare rocky planets and water worlds, we take the data provided by \cite{Zeng2019} and use the intermediate value of the radii for 300~K and 500~K at a given mass. For the envelopes, we numerically integrate convective structure with \citet{Saumon1995}'s adiabats for a 75\% H + 25\% He mixture up to 1~kbar. Above this we use the analytic solutions from \cite{Matsui1986} for radiative atmospheres with equilibrium temperature of 400~K \citep[see][for details]{Kurosaki2017}.   
    Since the mass measurement of TOI-2136b presents a large uncertainty, when compared to models, TOI-2136b could be a low-density planet ($\rho_{p} \sim 1.0 $ g cm$^{-3}$) with a significant gaseous envelope in its low mass limit, or a rocky planet with an Earth-like composition and a 1\% H$_2$ gaseous envelope in its upper mass limit. 

    To evaluate the potential for follow-up observations with the aim of detecting the presence of an atmosphere in TOI-2136b, we computed the transmission spectroscopy metric (TSM) of this planet following \cite{Kempton2018}. This metric is an estimate of the S/N achievable with the Near Infrared Imager and Slitless Spectrograph (NIRISS) instrument on the James Webb Space Telescope (\textit{JWST}, \citealp{Gardner2006}), using ten hours of observation time. 
For the median mass value of \hbox{TOI-2136b}, we find a TSM of $\sim 93$, and using the low and upper 1$\sigma$ mass limits, we find TSM values of 212 (for $M_{p\;low} = 2.04$ $\rm M_\oplus$) and 56 (for $M_{p\;up} = 7.75$ $\rm M_\oplus$). We compared the TSM obtained with the median mass value with those of previously discovered exoplanets that have a radius smaller than 4 $\rm R_\oplus$ (see Figure \ref{fig:MassRadius_TSM}, right panel). We find that TOI-2136b has a relatively large TSM when compared with other planets with radii around 2 $\rm R_\oplus$ and a similar $J$-band magnitude, and it also has an almost comparable TSM to hycean planets (labeled in Figure \ref{fig:MassRadius_TSM}, right panel), making it a very suitable target for JWST atmospheric exploration.


\section{Transmission spectroscopy} \label{sec:spectroscopy}

    TOI-2136b likely has a H/He gaseous envelope from the mass-radius relationship. Using the IRD high-resolution spectra taken during transit, it is possible to search for an extended atmosphere of H/He. In this particular case focusing on the helium infrared triplet at 10829.08, 10830.25, and 10830.34 \mbox{\AA} in air.

    Around the He absorption lines, there are $\rm OH^{-}$ emission lines and telluric $\rm H_{2}O$ absorption lines that sometimes contaminate a possible planetary absorption feature \citep{Nortmann+2018,Salz+2018,Alonso-Floriano+2019,Palle+2020}. To obtain the transmission spectrum, we corrected for these telluric absorption lines, as well as for the stellar RV shift and the planet RV. To obtain an $\rm OH^{-}$ emission model for our data, we fitted the strongest $\rm OH^{-}$ emission line with a Gaussian to estimate the central wavelength, and we compared it with previous works \citep{Palle+2020, Orell-Miquel+2022} to estimate the central wavelength of other $\rm OH^{-}$ emission lines. The region of the strongest emission was then masked.

\begin {figure} [htbp]
\begin{center}
 \includegraphics[width=0.5\textwidth] {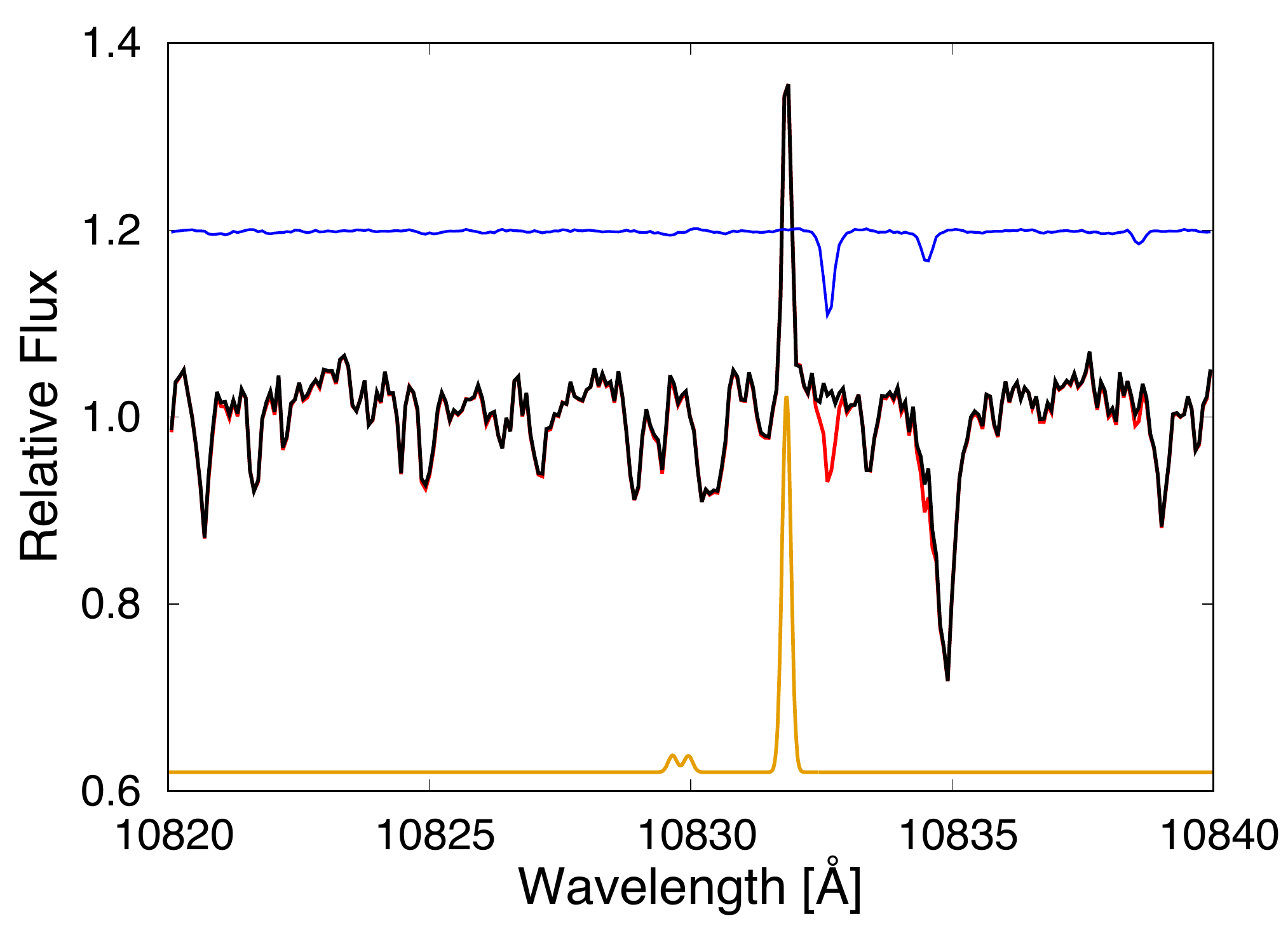}
\end{center}
\caption{Example of a single IRD spectrum of TOI-2136 around the helium triplet before (red) and after (black) the telluric correction. Blue and orange lines indicate the telluric absorption lines and $\rm OH^{-}$ emission lines. This figure indicates the position of the emission lines and the correction of the telluric absorption lines down to the noise level.}
\label{fig:telluric}
\end{figure}

    The telluric $\rm H_{2}O$ absorption lines were removed from the observed spectrum using a telluric standard star, and the correlation between line depth and airmass, following \citet{Kawauchi+2018}. We used a telluric standard star observed on 29 September 2020 as there were no observations of a telluric standard star on the same day. We created a telluric template spectrum by fitting the spectrum of the telluric standard star with the first spectrum of our transit time series. Fig~\ref{fig:telluric} shows that, with this correction, the telluric absorption lines are eliminated from the spectra down to the noise level.

\begin {figure*} [htbp]
\begin{center}
 \includegraphics[width=\textwidth] {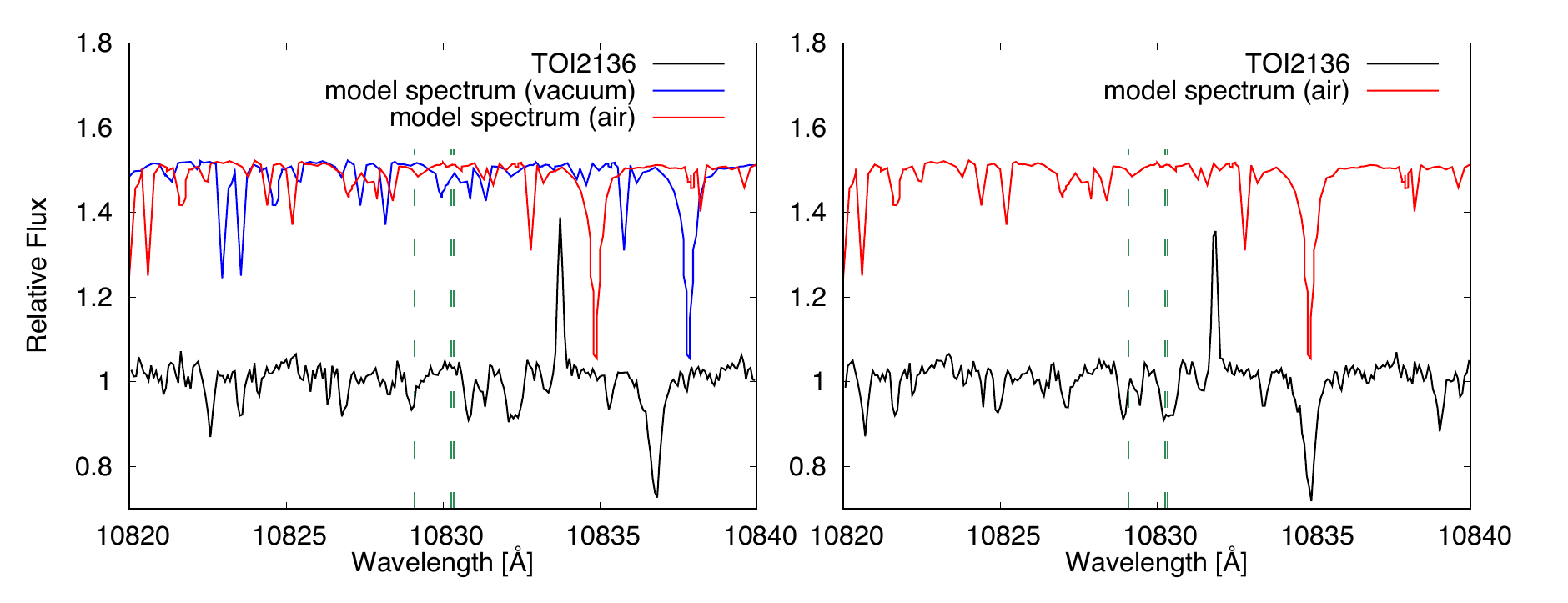}
\end{center}
\caption{Example of a single IRD spectrum of TOI-2136b (black) compared with the BT-Settl model \citep{Allard+2014} spectrum in vacuum (blue) and air (red). The left figure indicates the raw IRD spectrum and the right figure indicates the IRD-spectrum corrected shifted velocity. Vertical green dash lines in the left and right panels indicate the center of the predicted He triplet in air.}
\label{fig:stellarRV}
\end{figure*}

    The stellar system is shifted due to the Earth’s rotation and revolution. These stellar RVs in the barycentric reference frame were estimated from RA, Dec and observational time (UT) with \texttt{IRAF} task \texttt{rvcorrect}. We shifted the stellar spectra and corrected them (see the left panel of Figure~\ref{fig:stellarRV}). However, these spectra were not aligned with the model stellar spectra at air and vacuum wavelength.

    To correct the wavelength, we estimated the additional stellar RV shift from the model spectra with least-squares deconvolution (LSD) method \citep{Donati+1997}. Using this method, we created a line profile of the observed spectra in velocity space, convolved with the delta functions obtained from the center wavelength and the strength of atomic and molecular lines created in the model stellar atmosphere \citep{Watanabe+2020,Collier+2010}. We used the observed spectra in three orders around He lines (10681.55 - 10979.75 \mbox{\AA}), and the 426 absorption lines, assuming the stellar temperature and $\log g$ from VALD. 

    After we created a line profile of each frame, we fitted it with a flat line and a Gaussian to estimate the shifted velocity from the model. We obtained a value of -52.6 km\,s$^{-1}$ of the shifted velocity, and we shifted the observed spectrum by that value (see right panel of Figure~\ref{fig:stellarRV}). As a result, the observed spectra are consistent with the model spectra in air wavelength.

    After the above correction, we created the template stellar spectrum to combine all out-of-transit spectra (five frames) and divided each frame by it. 
    The planet signals were shifted in wavelength by planet orbital motion. We calculated the RV of the planet at each frame with Equation 8 of \citet{Khalafinejad+2017}. The RV of TOI-2136b changes from +1.7 km\,s$^{-1}$ to -1.5 km\,s$^{-1}$ during transit. We shifted each frame by the calculated velocity and combined all frames during transit to create the final transmission spectrum.

\begin {figure*} [htbp]
\begin{center}
\includegraphics[width=\textwidth] {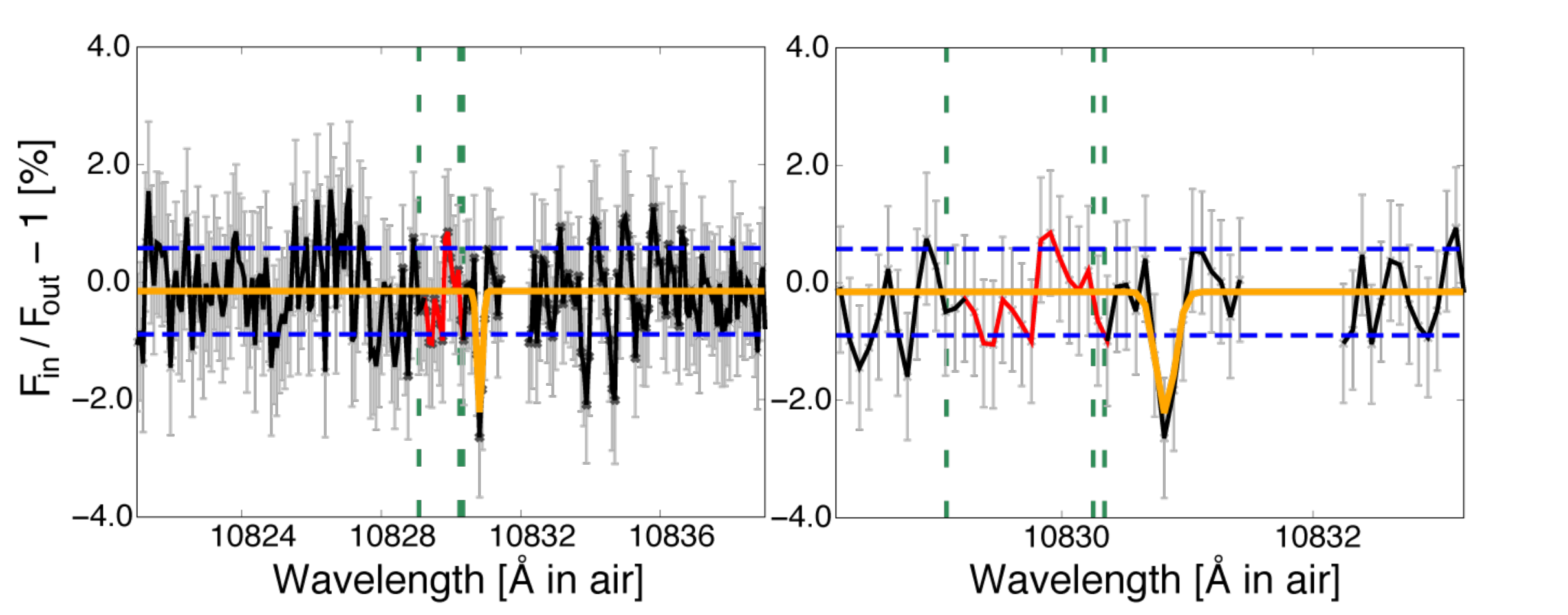}
\end{center}
\caption{Transmission spectrum of TOI-2136 around the He triplet. The blank region is where we mask the strongest $\rm OH^{-}$ emission. The red line is contaminated by the other shallower $\rm OH^{-}$ emission lines. The orange line is the best-fitted model with a Gaussian function. Vertical green dashed lines indicate the center of the predicted He triplet in air, and the horizontal blue lines are standard deviations between 10820 and 10840 \mbox{\AA}.}
\label{fig:transitratio}
\end{figure*}

    Figure~\ref{fig:transitratio} shows the final transmission spectrum of TOI-2136 over the \ion{He}{i}  triplet spectral region. A possible \ion{He}{i} signal at the triplet wavelengths (10829.081, 10830.25 and, 10830.34 \mbox{\AA}) is not apparent in the data, although some spectral features fall near the expected wavelengths.
    In the presence of an atmospheric outflow caused by stellar wind, the \ion{He}{i} absorption lines could be blue-shifted, but red-shifted helium absorption has been detected in HAT-P-32b \citep{Czesla+2022} and TOI 560.01 \citep{Zhang+2022}. Therefore, in order to estimate the detectability of \ion{He}{i}, we searched for the helium absorption line in a large range around the theoretical center of wavelength, and assumed that the largest absorption feature in this region of the spectra could be associated with the He triplet.
 
    To further explore this spectral feature at $\sim $10830.8 \mbox{\AA} (which is red-shifted by $\sim $ 15 km\,s$^{-1}$ from the expected wavelength of the strongest triplet line), we fitted a Gaussian with the central wavelength, depth, sigma, and baseline as free parameters, using the MCMC algorithm \texttt{emcee}. We used a normal prior for the central wavelength between 10829.21 and 10831.38 \mbox{\AA,} which correspond to about $\pm$ 30 km\,s$^{-1}$ from the center of the stronger \ion{He}{i} lines, and a uniform prior for sigma between 0 and 0.344, which corresponds to the thermal broadening at a temperature of 30,000 K.
    We obtained a statistically nonsignificant signal with a depth of 2.2 $\pm$ 1.1 \%, a sigma of 0.08 $\pm$ 0.06 and an EW of 4.3 $\pm$ 2.0 m\mbox{\AA} (Figure~\ref{fig:transitratio}). However, this spectral feature is only detected at the $2~\sigma$ level, and cannot be identified as significant or planetary in origin.
    Thus, our data only allow us to place upper limits to the presence of \ion{He}{i} in TOI-2136b's atmosphere. Any possible \ion{He}{i} signal would have an EW $<$ 7.8 m\mbox{\AA,} with 95\% confidence, and an absorption signal $<$ 1.44 \%, with 95\% confidence. 

\section{Discussion} \label{sec:dis}

\begin {figure} [htbp]
\begin{center}
 \includegraphics[width=0.5\textwidth] {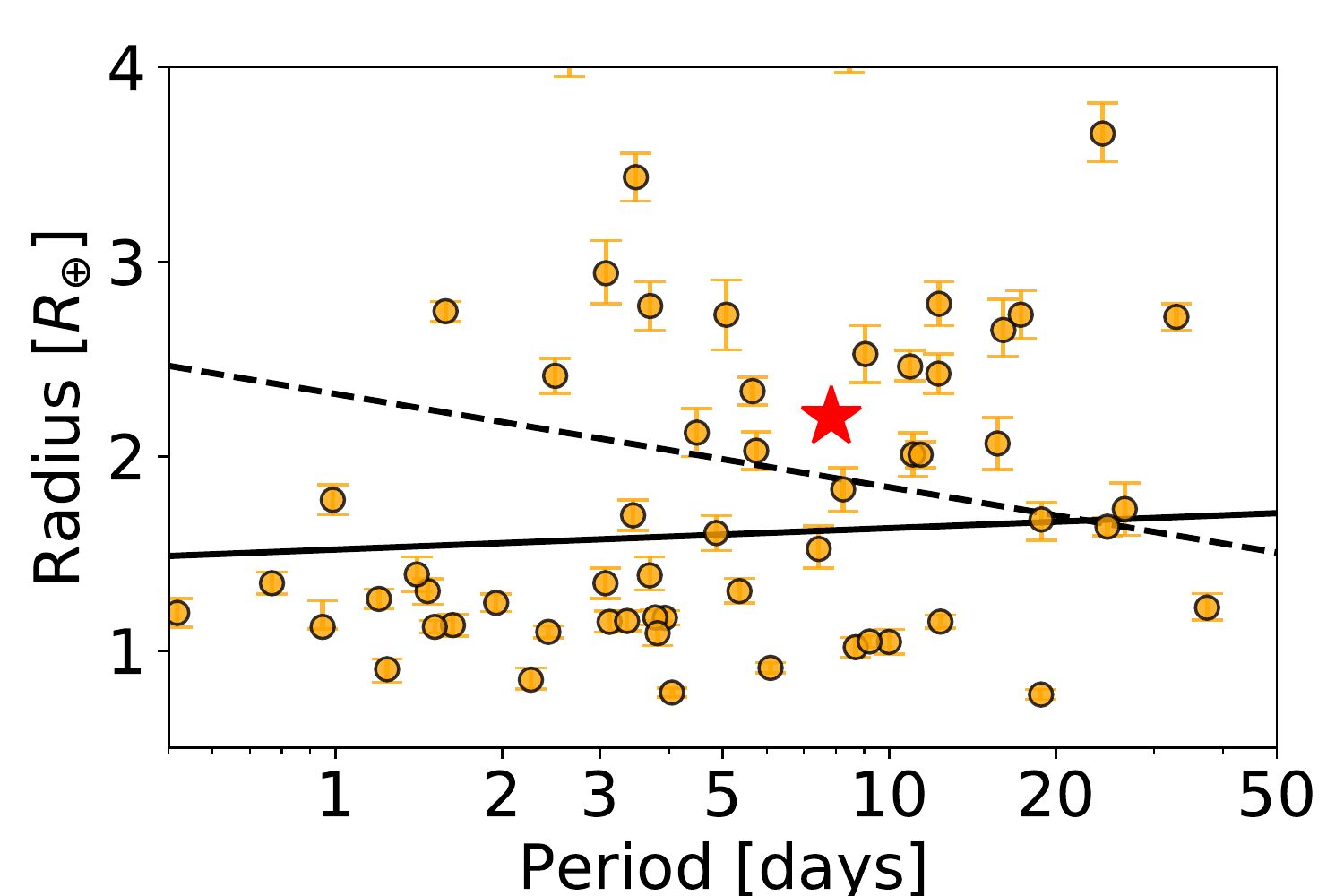}
\end{center}
\caption{Period-radius diagram for known exoplanets around M-type stars with radius measurements with a precision better than 8\% (orange points). The red star represents TOI-2136b. The black line and the dashed line indicate the radius valley around low-mass stars \citep{Cloutier+2020} and Sun-like stars \citep{Martinez+2019}, respectively.}
\label{fig:p-r}
\end{figure}

    In previous sections, we validated the planetary nature of TOI-2136.01. Figure \ref{fig:p-r} shows its location in a period-radius diagram, together with all other known planets around M-type stars whose radius measurements were better than 8$\%$ precise.
    In the figure, the black line indicates the radius valley measured around low-mass stars \citep{Cloutier+2020}, which is consistent with the gas-poor formation model \citep{Lopez&Rice2018}. The dashed black line indicates the radius valley measured around Sun-like stars \citep{Martinez+2019}, which is consistent with the photoevaporation \citep{Lopez&Rice2018} and core powered mass-loss models \citep{Gupta&Schlichting2019}. 
    TOI-2136b lies above both radius valleys, so it would be expected to be non-rocky in composition and posses some faction of volatile elements in its atmosphere. This prediction is consistent with our results in the mass-radius diagram (Figure~\ref{fig:MassRadius_TSM}). 

    Additionally, thanks to the multiple ground-based photometric observations, we obtained the radius of TOI-2136b with an uncertainty of $\sim$ 3\%. Presently, there are only a few planets measured with such high precision around M-type stars. Among them, 
    LTT3780c (2.42 $\pm$ 0.10 $\rm R_{\oplus}$, 6.29 $\pm$ 0.63 $\rm M_{\oplus}$, 2.88 $\pm$ 0.28 $\rm S_{\oplus}$; \citealt{Nowak+2020}), L321-32d (2.13 $\pm$ 0.06 $\rm R_{\oplus}$, 4.78 $\pm$ 0.43 $\rm M_{\oplus}$, 3.74 $\pm$ 0.61 $\rm S_{\oplus}$; \citealt{VanEylen+2021}), and TOI-776c (2.02 $\pm$ 0.14 $\rm R_{\oplus}$, 5.30 $\pm$ 1.80 $\rm M_{\oplus}$, 4.9 $\pm$ 0.2 $\rm S_{\oplus}$; \citealt{Luque+2021}) have similar parameters to TOI-2136b. 
    These planets are candidates to be hycean worlds, which are composed of water-rich interiors with massive oceans underlying H$_2$-rich atmospheres \citep{Madhusudhan+2021}.
    Hycean planets have received attention since a possible biosignature might be detectable in their atmospheres with a relatively modest amount of JWST observation time. 

 We also obtained a transmission spectrum of the planet around the \ion{He}{i} 10830 $\rm \AA$ absorption lines. In other works, helium absorption is detected on the atmosphere of three sub-Neptunes (TOI-560.01; \citealt{Zhang+2022}, GJ3470b; \citealt{Palle+2020}, and GJ1214b; \citealt{Orell-Miquel+2022}). To compare our results with previous \ion{He}{i} detections in the literature, we calculated the equivalent height of the atmosphere normalized by the scale height. This value for TOI-2136b is $<$ 114, adopting a mass of $M_{p}$=4.7 $\rm M_{\oplus}$ and is comparable with the value of TOI 560.01 (Figure 4 in \citealt{Orell-Miquel+2022}). TOI 560.01 has a detection of \ion{He}{i} (\citealp{Zhang+2022}), meaning that TOI-2136b likely has a less extended or a less rich H/He atmosphere than TOI 560.01. For GJ1214b and GJ3470b, the ratio between the equivalent height and atmospheric scale height is $\sim 57$ for both planets, about half of our upper limit for TOI-2136b. Despite our nondetection of \ion{He}{i} on this planet, the atmospheric metrics and its place in the mass-radius space suggest that TOI-2136b is an ideal candidate to search for atmospheric signals using ground-based high-resolution spectroscopic observations taken during a transit event and space-based observations with JWST.
    
    During the submission and revision stage of this manuscript, two studies on the TOI-2136 system were announced: \cite{Gan2022, Beard+2022}. \cite{Gan2022} used TESS data, ground-based follow-up observations, and the Spirou instrument on the Canada-France-Hawaii Telescope (CFHT) to obtain RV measurements; the study presents a planetary radius of $2.19 \pm 0.17$ $\rm R_{\oplus}$ and a mass of $6.4 \pm 2.4$ $\rm M_{\oplus}$ for the transiting planet. \cite{Beard+2022} used TESS data, ground-based follow-up observations, and the Hobby-Eberly Telescope (HET) instrument Habitable-Zone Planet Finder (HPF) instrument to obtain RV measurements. \cite{Beard+2022} presents a planetary radius of $2.09 \pm 0.08$ $\rm R_{\oplus}$ and a mass upper limit of $< 15$ $\rm M_{\oplus}$, with a median planet mass of $4.64$ $\rm M_{\oplus}$. Our results for the planetary mass and radius of TOI-2136b are consistent with both studies (taking into account the uncertainties). 
    
\section{Summary} \label{sec:sum}

    In this work we reported the follow-up efforts on the planet candidate TOI-2136b, a sub-Neptune planet orbiting around a nearby M3 dwarf. The initial observations made by TESS led to the detection of the transiting planet candidate. Multicolor, ground-based, follow-up photometric observations and RV measurements made it possible to confirm the planetary nature of the candidate and rule out false positives.

    Using the IRD instrument on the Subaru 8.2 m telescope, we obtained high-resolution spectra and derived stellar parameters for the planet host star. We find that TOI-2136 is a nearby ($d = 33.361 \pm 0.019 $ pc) M dwarf with $T_{eff}=3373 \pm 108$ K, $\left[\rm Fe/H\right] = 0.02 \pm 0.14$ dex, and an estimated stellar mass of $M_{*} = 0.3272 \pm 0.0082$ M$_\odot$ and a radius of $R_{*} = 0.3440 \pm 0.0099$ R$_\odot$. We also measured the rotation period of the star using MEarth and ZTF long-term photometric observations, and found a value of $P_{rot} = 82.56 \pm 0.45$ days. 

    We analyzed the available transit and RV observations of the planet candidate fitting the data jointly. 
    The transit observations and RV measurements from IRD were fitted simultaneously using an MCMC procedure that included Gaussian processes to model the systematic effects present in the TESS and RV measurements. We find that TOI-2136b has a planetary radius of $R_p = 2.20 \pm 0.07$ $\rm R_{\oplus}$ and a mass of $M_p = 4.7^{+3.1}_{-2.6}$ $\rm M_{\oplus}$ (with an upper limit on the mass value of $< 9.9$ $\rm M_{\oplus}$, with 95\% confidence). Our model supports a circular orbit ($e = 0.07^{+0.09}_{-0.05}$) with an orbital period of $7.851925 \pm 0.000016$ days. 

    Finally, we observed a transit event using high-resolution spectroscopy in an effort to detect the existence of an extended atmosphere around this planet. In particular we focused on the \ion{He}{i} 10830\mbox{\AA} absorption lines. We found no statistical significant extra absorption around the \ion{He}{i} lines that could be attributable to the planet, but we can place an upper limit on the presence of a He-rich extended atmosphere: an absorption depth of $<$ 1.44 \% and an EW of $<$ 7.8 m\mbox{\AA,} with 95\% confidence. 

    TOI-2136b is a small sub-Neptune planet orbiting a relatively bright ($J = 10.1$ mag) M star. Based on the TSM, this planet presents a very valuable opportunity for further atmospheric characterization studies and the search for additional planets in the system.

\begin{acknowledgements}
   This research is based on data collected at Subaru Telescope, which is operated by the National Astronomical Observatory of Japan.
    We are honored and grateful for the opportunity of observing the Universe from Maunakea, which has the cultural, historical and natural significance in Hawaii.
    The part of our data analysis was carried out on common use data analysis computer system at the Astronomy Data Center, ADC, of the National Astronomical Observatory of Japan. 
    This paper makes use of data from the MEarth Project, which is a collaboration between Harvard University and the Smithsonian Astrophysical Observatory. The MEarth Project acknowledges funding from the David and Lucile Packard Fellowship for Science and Engineering, the National Science Foundation under grants AST-0807690, AST-1109468, AST-1616624 and AST-1004488 (Alan T. Waterman Award), the National Aeronautics and Space Administration under Grant No. 80NSSC18K0476 issued through the XRP Program, and the John Templeton Foundation. Based on observations obtained with the Samuel Oschin 48-inch Telescope at the Palomar Observatory as part of the Zwicky Transient Facility project. 
    ZTF is supported by the National Science Foundation under Grant No. AST-1440341 and a collaboration including Caltech, IPAC, the Weizmann Institute for Science, the Oskar Klein Center at Stockholm University, the University of Maryland, the University of Washington, Deutsches Elektronen-Synchrotron and Humboldt University, Los Alamos National Laboratories, the TANGO Consortium of Taiwan, the University of Wisconsin at Milwaukee, and Lawrence Berkeley National Laboratories. Operations are conducted by COO, IPAC, and UW. 
    Our data reductions benefited from PyRAF and PyFITS that are the products of the Space Telescope Science Institute, which is operated by AURA for NASA.
    G. M. has received funding from the European Union's Horizon 2020 research and innovation programme under the Marie Sk\l{}odowska-Curie grant agreement No. 895525.
    R.L. acknowledges financial support from the Spanish Ministerio de Ciencia e Innovaci\'{o}n, through project PID2019-109522GB-C52, and the Centre of Excellence "Severo Ochoa" award to the Instituto de Astrof\'{i}sica de Andaluc\'{i}a (SEV-2017-0709).
    JK gratefully acknowledge the support of the Swedish National Space Agency (SNSA; DNR 2020-00104).
    This work is partly supported by JSPS KAKENHI Grant Numbers JP21K20376, JP21K13975, JP21H00035, JP21K20388, JP20K14518, JP20K14521, JP19K14783, JP18H05439, JP17H04574, by Grant-in-Aid for JSPS Fellows Grant Number JP20J21872, by JST CREST Grant Number JPMJCR1761, by Astrobiology Center PROJECT Research AB031014, by Astrobiology Center SATELLITE Research project AB022006, and the Astrobiology Center of National Institutes of Natural Sciences (NINS) (Grant Number AB031010).
    M.T. is supported by MEXT/JSPS KAKENHI grant Nos. 18H05442, 15H02063, and 22000005.
     Some of the observations in the paper made use of the High-Resolution Imaging instrument ‘Alopeke obtained under Gemini LLP Proposal Number: GN/S-2021A-LP-105. ‘Alopeke was funded by the NASA Exoplanet Exploration Program and built at the NASA Ames Research Center by Steve B. Howell, Nic Scott, Elliott P. Horch, and Emmett Quigley. Alopeke was mounted on the Gemini North (and/or South) telescope of the international Gemini Observatory, a program of NSF’s OIR Lab, which is managed by the Association of Universities for Research in Astronomy (AURA) under a cooperative agreement with the National Science Foundation. on behalf of the Gemini partnership: the National Science Foundation (United States), National Research Council (Canada), Agencia Nacional de Investigaci\'{o}n y Desarrollo (Chile), Ministerio de Ciencia, Tecnolog\'{i}a e Innovaci\'{o}n (Argentina), Minist\'{e}rio da Ci\^{e}ncia, Tecnologia, Inova\c{c}\~{o}es e Comunica\c{c}\~{o}es (Brazil), and Korea Astronomy and Space Science Institute (Republic of Korea).
\end{acknowledgements}

\bibliographystyle{aa} 
\bibliography{reference}

\begin{appendix} 
\section{Radial velocities}

\begin{table}
\caption{Subaru IRD RV measurements}\label{tab:RV_values}
\centering
\begin{tabular}{lcc}
\hline \hline
BJD  [-2459000] & RV  [ms$^{-1}$] & $\sigma_{RV}$  [ms$^{-1}$] \\
\hline
119.71575590 & 0.440 & 4.250 \\
119.72649620 & 3.360 & 4.980 \\
119.73722720 & -4.750 & 4.530 \\
119.74796610 & -9.740 & 4.260 \\
119.75869830 & -2.490 & 4.520 \\
119.76943850 & 1.920 & 4.710 \\
119.78016590 & -8.130 & 4.720 \\
119.79090390 & -11.730 & 4.680 \\
119.80164520 & -0.580 & 4.560 \\
119.81237620 & -5.750 & 4.290 \\
119.82310590 & -8.760 & 4.140 \\
156.80336260 & 8.220 & 5.190 \\
189.68940840 & 3.270 & 5.460 \\
248.16697680 & 10.540 & 5.990 \\
276.14592460 & 2.520 & 8.070 \\
276.16016820 & -3.010 & 5.390 \\
337.91216630 & 18.250 & 5.140 \\
338.91002330 & 8.150 & 3.480 \\
372.02252050 & 11.040 & 3.950 \\
372.03322450 & -2.960 & 3.930 \\
373.86437420 & 2.500 & 3.670 \\
389.85266460 & -10.380 & 4.380 \\
390.83389110 & 1.860 & 4.170 \\
456.79522350 & 2.680 & 4.300 \\
456.80944320 & -1.970 & 4.290 \\
466.85547440 & 3.760 & 3.960 \\
466.86970380 & 2.520 & 3.880 \\
487.82708040 & -9.330 & 3.970 \\
487.84130740 & -2.180 & 4.050 \\
487.85553810 & -5.660 & 3.910 \\
502.80575910 & -2.400 & 3.790 \\
502.81998640 & 1.810 & 3.770 \\
507.80134910 & -0.340 & 4.050 \\
507.81558570 & -8.810 & 4.010 \\
509.77572120 & -3.420 & 4.220 \\
509.78996010 & -4.860 & 4.230 \\
514.69360020 & -14.970 & 3.980 \\
514.70783790 & -16.860 & 3.970 \\
\hline
\end{tabular}
\end{table}

\section{Stellar rotation priors} 

\begin{table}
\caption{Stellar rotation period fitted parameters, prior functions, and limits.}\label{tab:Prot_priors}
\centering
\begin{tabular}{lc}
\hline \hline
Parameter & Value \\
\hline
$\log B$ & $\mathcal{U}(-12,6)$ \\
$\log C$ & $\mathcal{U}(-12, 12)$ \\
$\log L$  & $\mathcal{U}(-12, 12)$ \\
$\log P_{rot}$ &  $\mathcal{N}(1.6,6.0)$ \\
$\log_{jitter}$ & $\mathcal{U}(10^{-6},10)$ \\
\hline
\end{tabular}
\tablefoot{$\mathcal{U}$, $\mathcal{N}$ represent Uniform and Normal prior functions, respectively.}
\end{table}

\section{Light curve and radial velocity joint fit}

\begin{table}
\caption{Global fit parameters, prior functions, and limits.}\label{tab:pparam_priors}
\centering
\begin{tabular}{lc}
\hline \hline
Parameter & Value \\
\hline
\multicolumn{2}{c}{(Fitted orbital and transit parameters priors)} \\
$R_{p}/R_{*}$ & $\mathcal{U}(0.005,0.35)$ \\
$T_{c}$  [BJD] & $\mathcal{U}(2459212.5, 2459215.5)$ \\
$P$ [days] & $\mathcal{U}(3.8, 11.8)$ \\
$\rho_*$ [g cm$^{-3}$] &  $\mathcal{N}(11.33,3.06)$ \\
$b = (a/R_\star) \cos(i) \left( \frac{1-e^2}{1+e\sin(\omega)} \right)$ & $\mathcal{U}(0,1)$ \\
$\sqrt{e}\cos(\omega)$ & $\mathcal{U}(-1,1)$ \\
$\sqrt{e}\sin(\omega)$ & $\mathcal{U}(-1,1)$ \\
$\gamma_0 - \langle \gamma_0 \rangle$ [m/s] & $\mathcal{U}(-42.075, 37.925)$ \\
$K$ [m/s] & $\mathcal{U}(0,110)$ \\
$\sigma_{RV}$ [m/s] & $\mathcal{U}(0,10)$ \\
\multicolumn{2}{c}{(LD coefficients priors)} \\
$q_1 = (u_1 + u_2)^2$ & $\mathcal{U}(0,1)$  \\
$q_2 = 0.5u_1/(u_1+u_2)$ & $\mathcal{U}(0,1)$  \\ 
\multicolumn{2}{c}{(Fitted GP parameters priors)} \\
$\log(c_1)$ TESS S26 & $\mathcal{U}(-8,6)$ \\
$\log(\tau_1)$ TESS S26 & $\mathcal{U}(-2.65,6)$ \\
$\log(c_1)$ TESS S40 & $\mathcal{U}(-8,6)$ \\
$\log(\tau_1)$ TESS S40 & $\mathcal{U}(-2.65,6)$ \\
$c_2$ IRD & $\mathcal{U}(0,100)$ \\
$\tau_2$ IRD & $\mathcal{U}(10^{-3},150)$ \\
\hline
\end{tabular}
\tablefoot{$\mathcal{U}$, $\mathcal{N}$ represent Uniform and Normal prior functions, respectively.}
\end{table}

\begin{figure*}
   \centering
   \includegraphics[width=\textwidth]{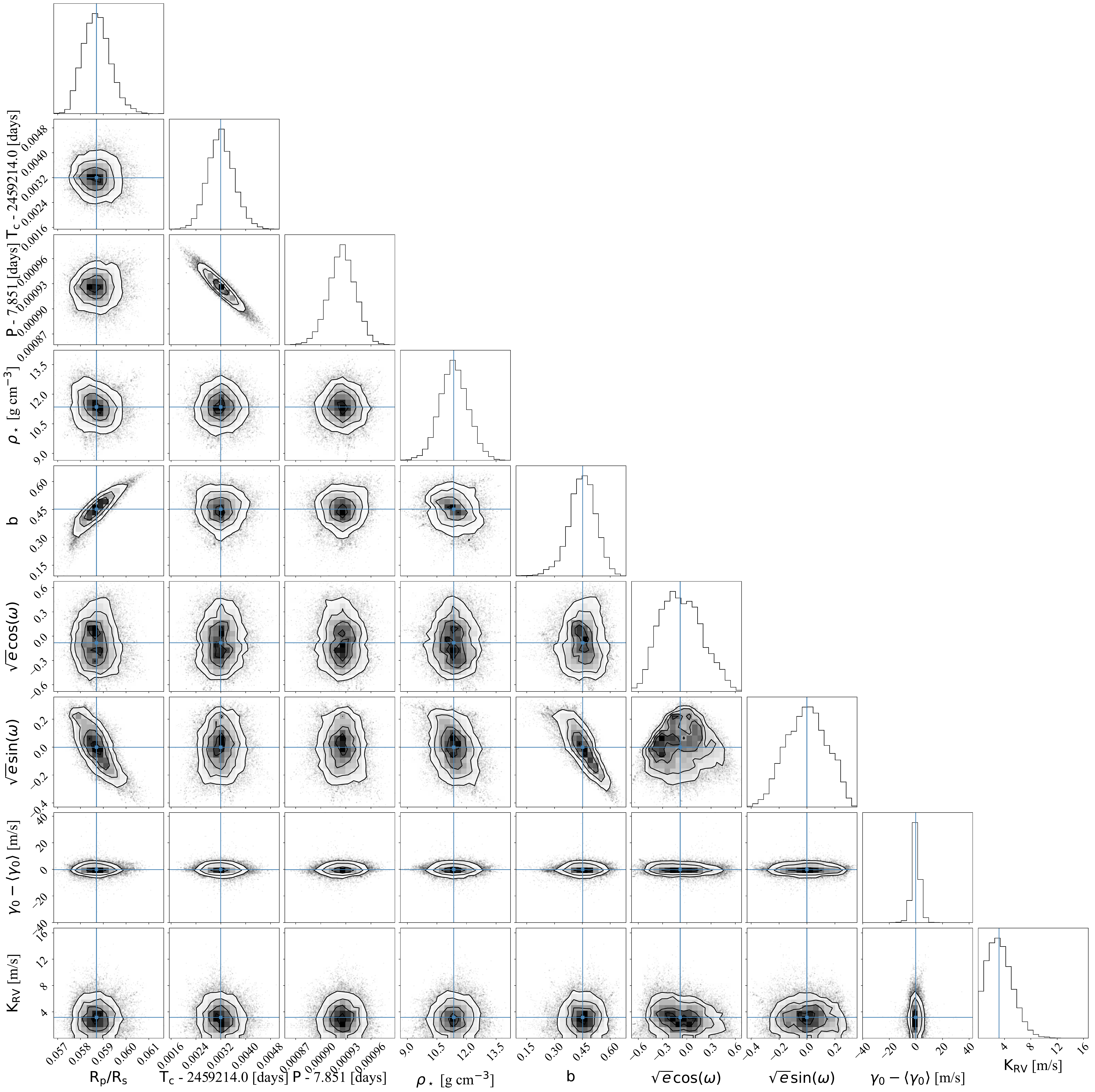}
   \caption{Correlation plot for the fitted orbital parameters. Limb-darkening coefficients and systematic effect parameters were intentionally left out for easy viewing. The blue lines give the median values of the parameters.}
    \label{Fig:Fit_ParamDistr_CornerPlot}
\end{figure*}

\begin {figure*} [htbp]
\begin{center}
 \includegraphics[width=\textwidth] {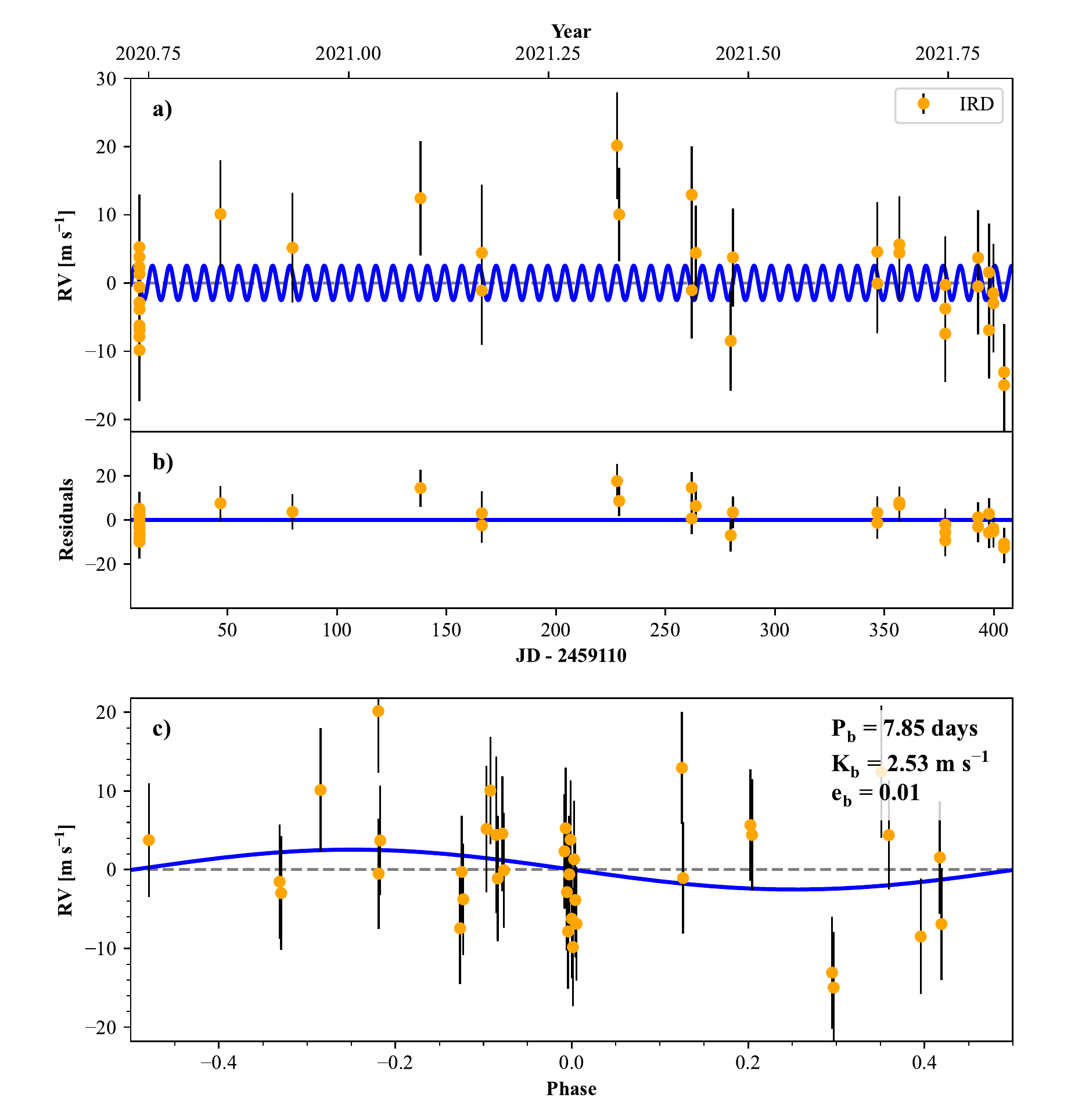}
\end{center}
\caption{RV measurements of TOI-2136 taken with IRD. The top panel presents the time series and best fitting model without GPs. The middle panel presents the residuals of the fit. The bottom panel shows the RV measurements in phase after subtracting the red noise.}
\label{fig:Subaru_RV_plot_withoutGPs}
\end{figure*}

\begin{figure*}
   \centering
   \includegraphics[width=\textwidth]{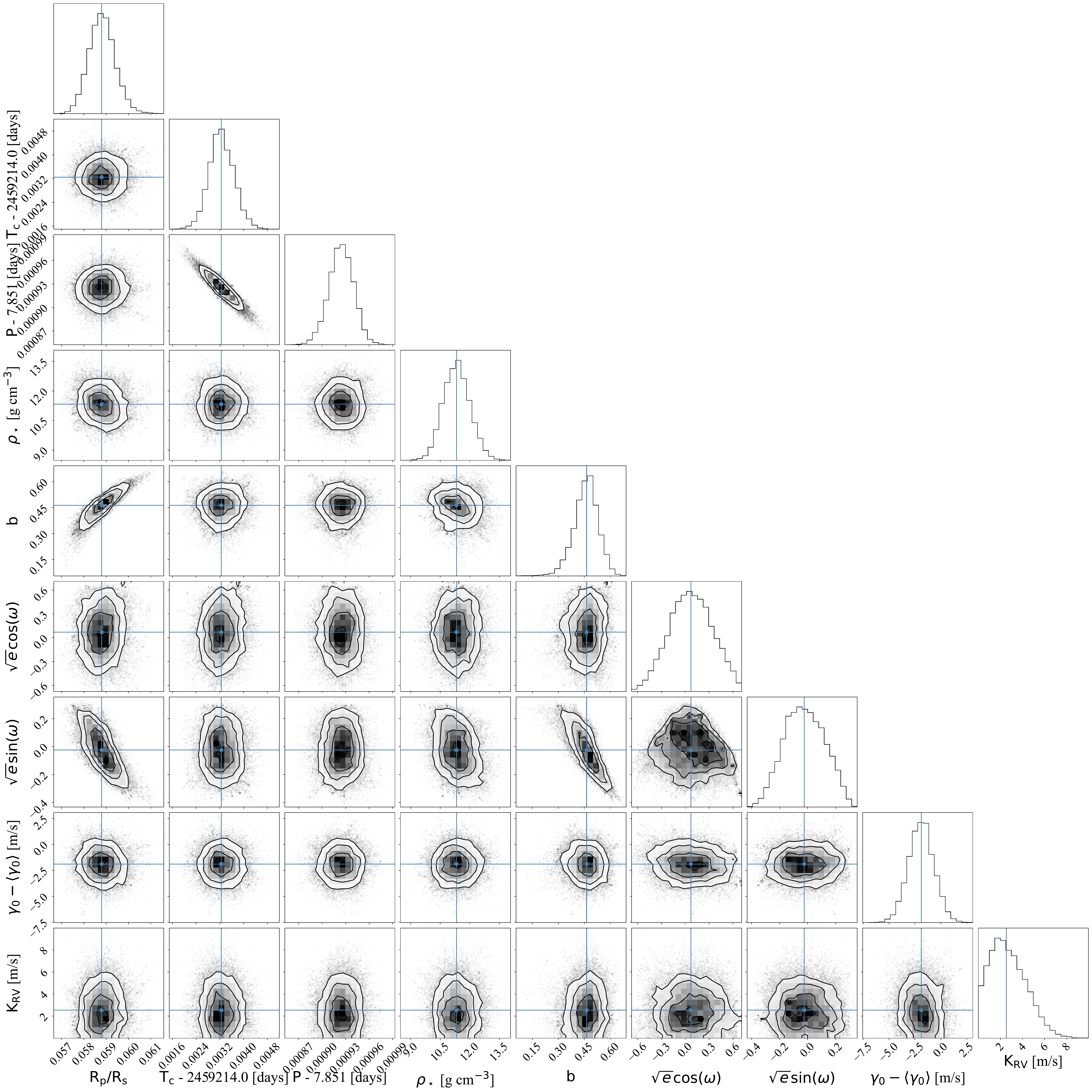}
   \caption{Correlation plot for the fitted orbital parameters without GPs. Limb-darkening coefficients and systematic effect parameters were intentionally left out for easy viewing. The blue lines give the median values of the parameters.}
    \label{Fig:Fit_ParamDistr_CornerPlot_withoutGPs}
\end{figure*}

\end{appendix}

\end{document}